\documentclass[preprint,12pt]{elsarticle}

\usepackage{graphicx}

\usepackage{amsmath,amssymb,amsfonts}

\def\be{\begin{equation}}
\def\ee{\end{equation}}
\def\ba{\begin{eqnarray}}
\def\ea{\end{eqnarray}}
\def\bd{\begin{displaymath}}
\def\ed{\end{displaymath}}
\def\bq{\begin{eqnarray}}
\def\eq{\end{eqnarray}}

\journal{Annals of Physics}

\begin{document}

\begin{frontmatter}




\title{Understanding interference experiments with polarized light
through photon trajectories}


\author[label1]{A. S. Sanz} \ead{asanz@imaff.cfmac.csic.es}
\author[label2]{M. Davidovi\'c}
\author[label3]{M. Bo\v zi\'c}
\author[label1]{S. Miret-Art\'es}

\address[label1]{Instituto de F\'{\i}sica Fundamental,
Consejo Superior de Investigaciones Cient\'{\i}ficas,
Serrano 123, 28006 Madrid, Spain}

\address[label2]{Faculty of Civil Engineering, University of Belgrade,
Bulevar Kralja Aleksandra 73, 11000 Belgrade, Serbia}

\address[label3]{Institute of Physics, University of Belgrade,
Pregrevica 118, 11080 Belgrade, Serbia}

\begin{abstract}
Bohmian mechanics allows to visualize and understand the
quantum-mecha\-ni\-cal behavior of massive particles in terms
of trajectories.
As shown by Bialynicki-Birula, Electromagnetism also admits a
hydrodynamical formulation when the existence of a wave function
for photons (properly defined) is assumed.
This formulation thus provides an alternative interpretation of
optical phenomena in terms of photon trajectories, whose flow yields a
pictorial view of the evolution of the electromagnetic energy density
in configuration space.
This trajectory-based theoretical framework is considered here to study
and analyze the outcome from Young-type diffraction experiments within
the context of the Arago-Fresnel laws.
More specifically, photon trajectories in the region behind the two
slits are obtained in the case where the slits are illuminated by a
polarized monochromatic plane wave.
Expressions to determine electromagnetic energy flow lines and photon
trajectories within this scenario are provided, as well as a procedure
to compute them in the particular case of gratings totally transparent
inside the slits and completely absorbing outside them.
As is shown, the electromagnetic energy flow lines obtained allow
to monitor at each point of space the behavior of the electromagnetic
energy flow and, therefore, to evaluate the effects caused on it by
the presence (right behind each slit) of polarizers with the same or
different polarization axes.
This leads to a trajectory-based picture of the Arago-Fresnel laws for
the interference of polarized light.
\end{abstract}


\begin{keyword}
Maxwell equations \sep
Photon wave function \sep
Hydrodynamic formulation of Electromagnetism \sep
Electromagnetic energy flow line \sep
Photon trajectory \sep
Bohmian mechanics \sep
Quantum trajectory

\PACS
03.50.De 
\sep
42.50.-p 
\sep
42.50.Ar 
\sep
03.65.Ta 
\sep
42.25.Hz 
\sep
42.25.Ja 

\end{keyword}

\end{frontmatter}



\section{\label{sec1} Introduction}

By means of different experimental techniques based on the use of
low-intensity beams, nowadays it is possible to observe the gradual
emergence of interference patterns in the case of both massive
particles \cite{tonomura,shimizu} and light (photons) \cite{parker,%
weis,ophe}.
That is, the patterns are built up by detecting single events (one
by one) which can be followed on a screen as a function of time.
In the pattern formation sequence, at short time scales, when the
number of events (or detections) recorded is still relatively low,
seemingly random distribution of points can be observed on a screen
used to monitor the experiment.
However, as time proceeds and the number of detections increases
sufficiently, the well-known band structure of alternate light and
dark interference fringes starts to appear.
Hence, moving from the darker regions of the pattern to the lighter
ones means that the density of particles detected increases.
This kind of experiments constitutes a nice manifestation of the
statistical nature of Quantum Mechanics, which, in the large
count-number limit, establishes that both massive particles \cite{born}
and photons \cite{loudon} distribute according to a certain
probability density.

Based on this kind of experiments, one could assume that each point on
the screen corresponds to the final position of the trajectory pursued
by a detected particle.
This immediately leads to considering trajectory-based formulations to
explain interference processes, where the analysis of the properties
displayed by (particle) trajectories provides a dynamical explanation
and interpretation of the process.
In other words, these formulations would contain the mechanisms
that allow to account for the appearance of fringes in interference
experiments with quantum particles (either massive particles or
photons) in the large-number limit of single-event counts, this outcome
being in agreement with the results obtained from standard wave-based
formulations.
This is an important and remarkable advantage, for these approaches
thus constitute physical models where both corpuscular and wave-like
properties are compatible.

In the case of interference experiments with massive particles,
different trajectory-based studies can be found in the literature
\cite{philippidis,sanz1a,sanz1b,sanz2a,sanz2b,gondran,sanz3,mirjana2,%
mirjana1}, which are aimed to establish a bridge between the single
counts observed experimentally and the statistical quantum-mechanical
description in terms of smooth, continuous probability densities.
Many of them rely on {\it Bohmian mechanics} \cite{madelung,bohm,%
takabayashi,holland1,detlef}, a hydrodynamic formulation of Quantum
Mechanics in terms of trajectories, which, to some extent, can be
considered at the same level as Newtonian mechanics with respect to
(classical) Liouvillian mechanics.
Bohmian trajectories evolve following the streamlines associated with
the quantum probability current density (quantum flow) and, therefore,
reproduce {\it exactly} the quantum-mechanical probability distribution
in both the near and the far field when a large number of them is
considered, in particular, in interference experiments
\cite{sanz1a,sanz1b,sanz3}.
Alternatively, the emergence of interference patterns in the far field
has also been simulated by using trajectories determined from the
momentum distribution (MD trajectories) associated with the particle
wave function \cite{mirjana1}.

In the case of interference processes with light (photons), similar
descriptions are not so numerous.
Some early attempts to explain interference and diffraction in terms
of electromagnetic energy (EME) flow lines were those of Braunbek and
Laukien \cite{braunbek} and Prosser \cite{prosser1,prosser2}.
Within this approach, based on Maxwell's equations, the photon
trajectories are associated with the EME flow lines \cite{prosser2},
which are obtained after solving a trajectory equation arising from
the {\it Poynting energy flow vector} \cite{jackson}.
Similarly, but starting from the Dirac equation instead of Maxwell's
ones, Ghose {\it et al.}\ \cite{ghose} also determined photon
trajectories.
The formal grounds of these approaches are similar to those of
the hydrodynamic formulation of Electromagnetism developed by
Bialynicki-Birula \cite{bialynicki1,bialynicki2,bialynicki3,%
bialynicki4}.
This formalism arises after assuming that, as happens with massive
particles, photons can also be described by a well-defined wave
function, this being a problem which has received much attention in the
literature \cite{bialynicki1,bialynicki2,bialynicki3,bialynicki4,sipe,%
landau,dirac,cook,inagaki,scully,kobe,berry,holland2,smith-raymer1,%
smith-raymer2,yong}.
Specifically, by following the same argumentation that led Dirac
to the relativistic equation for the electron, Bialynicki-Birula
\cite{bialynicki1,bialynicki2,bialynicki3,bialynicki4} reaches
Maxwell's equations.
Accordingly, though these equations have a purely classical origin,
they can be expressed as a Dirac-like equation, with the wave function
satisfying it being a complex vector encompassing both the electric
and magnetic field, namely the Riemann-Silberstein \cite{riemann,%
bateman} (see Appendix~\ref{appA}).

More specifically, in order to better understand the relationship
between EME flow lines and photons, consider the following.
As mentioned above, the EME flow lines describe the paths along which
the EME flows.
Assuming that a wave function defined in terms of the EM field can
be associated to a quantum of energy or photon \cite{bialynicki1,%
bialynicki2,bialynicki3,bialynicki4,sipe,landau,dirac,cook,inagaki,%
scully,kobe,berry,holland2,smith-raymer1,smith-raymer2,yong}, the
probability that a photon arrives at a certain region on a screen
will be proportional to the EME density at that region, i.e., to
the density of flow lines reaching such a region.
Hence, one could interpret that the trajectory arrivals somehow
simulate the arrivals of single photons to the screen, as seen
experimentally \cite{weis,ophe}.
That is, the trajectories are the paths along which the EME flows and
ensembles of them will constitute the eventual/possible paths taken by
single photons.
However, a photon itself has to be represented by an ensemble of
trajectories in order to satisfy the condition of energy quantization.
Throughout this work, we are going to use the term ``photon
trajectory'' to denote any of the eventual EME flow lines or
trajectories that a photon can pursue, but taking into account the
previous discussion.

Recently, we have considered \cite{asanz-phys-scr} Prosser's
approach (a simplified view of the more general Bialynicki-Birula
hydrodynamical formulation) to study Young's experiment and the Talbot
effect for a monochromatic, linearly polarized incident
electromagnetic (EM) field (or wave) reaching the diffracting grating.
This study was carried out in the region between the grating (behind
it) and the detection screen, thus the photon trajectories were
computed directly from the diffracted wave function, which was
obtained by requiring that the components of both the electric and the
magnetic vector fields satisfied Maxwell's equations as well as the
boundary conditions imposed by the grating.
The trajectories obtained showed how the EME redistributes in space
from the grating to the detection screen, located far away from the
former in the Fraunhofer region \cite{sanz1a,sanz1b}.
By properly sampling the field at the slits, one could also observe
how, as time proceeds, the accumulation of trajectory arrivals, which
eventually led to the appearance of the well-known fringe-like pattern,
just as in the analogous real experiment \cite{parker,weis,ophe} or as
in Quantum Mechanics, where the formation of similar patterns arises
after a large count of particle trajectories.

In this work, we go a step forward and present a more general
trajectory-based analysis of Young-type experiments carried out
with polarized light.
Experiments with high-intensity polarized beams are well-known since
the beginning of the XIXth century, when Arago and Fresnel enunciated
their laws for Young's interference experiment with polarized light
\cite{barakat}.
According to these laws, no interference pattern is observed if,
for example, the two interfering beams are linearly polarized in
orthogonal directions \cite{hunt} or both are elliptically polarized
but with opposite handedness \cite{pescetti}.
By means of the Bialynicki-Birula hydrodynamical formulation of
Electromagnetism \cite{bialynicki1,bialynicki2,bialynicki3,bialynicki4},
here we provide a trajectory picture for this kind of experiments.
Since we are interested in the interference phenomenon and its
relationship with the polarization of the interfering waves, we start
the calculation of the photon trajectories behind the slits.
By assuming that the incoming wave is a polarized monochromatic EM
plane wave, the components of the electric and magnetic fields can be
expressed in terms of a function that explicitly takes into account
the boundary conditions imposed by the two slits (here, the grating is
considered to be totally transparent inside the slits and completely
absorbing outside them).
Within this scenario, we provide expressions to determine the photon
trajectories as well as a procedure to compute them in the particular
case of gratings totally transparent inside the slits and completely
absorbing outside them.
As is shown, the photon trajectories obtained allow to monitor at each
point of space the behavior of the electromagnetic energy flow and,
therefore, to evaluate the effects caused on it by the presence (right
behind each slit) of polarizers with the same or different polarization
axes.
This thus constitutes a very interesting pictorial view of the
Arago-Fresnel laws for the interference of polarized light, consistent
with the more conventional one based on standard Electromagnetism.

The organization of this paper is as follows.
In order to be self-contained, in Section~\ref{sec2} we introduce
some fundamental theoretical grounds related to the photon-trajectory
formalism, starting from Maxwell's equation, as also done within the
framework of the Bialynicki-Birula hydrodynamical formulation of
Electromagnetism.
Particularly, we will focus on those elements related to the analysis
of experiments with polarized light.
In Section~\ref{sec3}, we describe the incident EM field and its
polarization, as well as the derivation of the corresponding photon
trajectories before reaching the interference grating.
In Section~\ref{sec4}, the EM field at and behind the grating is
obtained as a solution of Maxwell's equations.
This field is going to be used to calculate later the associated photon
trajectories.
In Section~\ref{sec5}, we present and discuss the photon trajectories
behind a two-slit grating in the case of an incident circularly
polarized EM field.
In Section~\ref{sec6} an analogous analysis is carried out, but when
the slits are covered with polarizers with orthogonal polarization
axes.
Calculations of photon trajectories for the cases of incident linearly
and circularly polarized EM fields are shown.
Finally, in Section~\ref{sec7} the conclusions arisen from this work
are summarized.


\section{\label{sec2} Maxwell's equations, EME flow lines and photon
trajectories}

Consider a monochromatic EM field propagating in vacuum.
Under these conditions, the electric and magnetic components of this
field can be expressed as harmonic waves, as
\begin{equation}
 \tilde{\bf E}({\bf r},t) = {\bf E}({\bf r}) e^{-i \omega t} ,
 \qquad
 \tilde{\bf H}({\bf r},t) = {\bf H}({\bf r}) e^{-i \omega t} ,
 \label{eq2}
\end{equation}
respectively, where the space dependent parts of these fields satisfy
the time-independent Maxwell's equations,
\begin{eqnarray}
 \nabla \cdot {\bf E}({\bf r}) & = & 0 ,
 \label{eq3} \\
 \nabla \cdot {\bf H}({\bf r}) & = & 0 ,
 \label{eq4} \\
 \nabla \times {\bf E}({\bf r}) & = & i \omega \mu_0 {\bf H}({\bf r}) ,
 \label{eq5} \\
 \nabla \times {\bf H}({\bf r}) & = &
   - i \omega \epsilon_0 {\bf E}({\bf r}) ,
 \label{eq6}
\end{eqnarray}
as well as the boundary conditions associated with the particular
problem under study.
Equivalently, from (\ref{eq3})-(\ref{eq6}), it is readily shown that
both ${\bf E}({\bf r})$ and ${\bf H}({\bf r})$ satisfy the Helmholtz
equation,
\begin{eqnarray}
 \nabla^2 {\bf E}({\bf r}) + k^2 {\bf E}({\bf r}) = 0 ,
 \label{eq7} \\
 \nabla^2 {\bf H}({\bf r}) + k^2 {\bf H}({\bf r}) = 0 ,
 \label{eq8}
\end{eqnarray}
where $k = \omega/c$.

The EME flow lines are obtained from the real part of the time-averaged
complex Poynting vector \cite{jackson},
\begin{equation}
 {\bf S}({\bf r}) = \frac{1}{2}\
  {\rm Re} [ {\bf E}({\bf r}) \times {\bf H}^*({\bf r}) ] ,
 \label{eq9}
\end{equation}
as follows.
We know that this vector describes the flow of the time-averaged EME
density through space,
\begin{equation}
 U({\bf r}) = \frac{1}{4} \left[
  \epsilon_0 {\bf E}({\bf r}) \cdot {\bf E}^*({\bf r})
  + \mu_0 {\bf H}({\bf r}) \cdot {\bf H}^*({\bf r}) \right] .
 \label{eq45}
\end{equation}
That is, the EME density is transported through space as a flow
described by ${\bf S}({\bf r})$.
Formally, this can be expressed as
\be
 {\bf S}({\bf r}) = U({\bf r}) {\bf v} ,
 \label{eq44bb}
\ee
where ${\bf v}$ is a local effective velocity vector field, namely the
{\it ray velocity} \cite{bornwolf}.
This velocity indicates the direction of the EME flow at each space
point and its magnitude is equal to the EME that crosses in unit time
an area perpendicular to the flow direction (given by the Poynting
vector ${\bf S}({\bf r})$) divided by the EME per unit volume.
Taking into account the Bialynicki-Birula hydrodynamical viewpoint,
one can assume that the EME travels along streamlines defined by
(\ref{eq44bb}).
Therefore, this equation can be recast in a more convenient way as
\begin{equation}
 \frac{d{\bf r}}{ds} = \frac{1}{c}
  \frac{{\bf S}({\bf r})}{U({\bf r})} ,
 \label{eq44}
\end{equation}
whose solutions, ${\bf r}(s)$, are the streamlines or EME flow
lines followed by the EME density in configuration space or, within a
Bohmian-like reinterpretation of the Bialynicki-Birula hydrodynamical
formulation, the photon trajectories.
Here, $s$ is simply a parameter which labels the evolution across space
of the corresponding trajectory, however, for practical purposes, later
we will reparametrize the solutions in terms of one of the coordinates
---in particular, we consider the reparametrization $\{x,y,z\}(s) \to
\{y,z\}(x)$.

In the particular case of interference experiments, a functional form
for (\ref{eq44}) can be found by means of the following considerations.
The screen containing the grating is on the $XZ$ plane, at $y = 0$, and
the slits are parallel to the $z$-axis, their width ($\delta$) being
much larger along this direction than along the $x$-direction (i.e.,
$\delta_z \gg \delta_x$).
Accordingly, we can assume that the EME density is independent of the
$z$-coordinate and, therefore, the electric and magnetic fields will
not depend either on this coordinate.
This allows us to consider a simplification in the analytical
treatment, for, as mentioned in \cite{bornwolf}, a problem independent
of one Cartesian coordinate is essentially scalar and, therefore, can
be formulated in terms of one dependent variable.
Thus, if we introduce here
\begin{equation}
 \frac{\partial {\bf H}}{\partial z} = {\bf 0} =
 \frac{\partial {\bf E}}{\partial z}
 \label{eq12}
\end{equation}
into Eqs.~(\ref{eq5}) and (\ref{eq6}), we obtain two independent sets
of equations,
\begin{eqnarray}
 \frac{\partial E_z}{\partial y} & = &
  \frac{i \omega}{\epsilon_0 c^2} H_x , \qquad
 \frac{\partial E_z}{\partial x} =
  - \frac{i \omega}{\epsilon_0 c^2} H_y , \nonumber \\
 \frac{\partial H_y}{\partial x} & - & \frac{\partial H_x}{\partial y}
  = - i \omega \epsilon_0 E_z ,
 \label{eq13}
\end{eqnarray}
and
\begin{eqnarray}
 \frac{\partial H_z}{\partial y} & = &
  - i \omega \epsilon_0 E_x , \qquad
 \frac{\partial H_z}{\partial x} =
  i \omega \epsilon_0 E_y , \nonumber \\
 \frac{\partial E_y}{\partial x} & - & \frac{\partial E_x}{\partial y}
  = \frac{i \omega}{\epsilon_0 c^2} H_z .
 \label{eq14}
\end{eqnarray}
The set (\ref{eq13}) only involves $H_x$, $H_y$ and $E_z$, and
therefore is commonly referred as a case of $E$-polarization, while
the set (\ref{eq14}), which only involves $E_x$, $E_y$ and $H_z$, is
referred to as $H$-polarization.

More specifically, as infers from the set of equations (\ref{eq13}), in
the case of $E$-polarization the electric field is polarized along the
$z$-direction, while the magnetic field is confined to the plane $XY$.
That is, $E_{e,x} = E_{e,y} = H_{e,z} = 0$, with the components of the
magnetic field satisfying
\begin{equation}
 H_{e,x} = - \frac{i \epsilon_0 c^2}{\omega}
  \frac{\partial E_{e,z}}{\partial y} , \qquad
 H_{e,y} = \frac{i \epsilon_0 c^2}{\omega}
  \frac{\partial E_{e,z}}{\partial x} .
 \label{eq16}
\end{equation}
Substituting these expressions for $H_x$ and $H_y$ into the second line
of (\ref{eq13}) yields
\begin{equation}
 \frac{\partial^2 E_{e,z}}{\partial x^2}
  + \frac{\partial^2 E_{e,z}}{\partial y^2} + k^2 E_{e,z} = 0 .
 \label{eq17}
\end{equation}
We thus have
\begin{eqnarray}
 {\bf E}_e & = & E_{e,z} \hat{\bf z} ,
 \label{eq21} \\
 {\bf H}_e & = &   H_{e,x} \hat{\bf x} + H_{e,y} \hat{\bf y} ,
 \label{eq22}
\end{eqnarray}
with $E_{e,z}$ satisfying Helmholtz's equation, according to
(\ref{eq17}).
Analogously, in the case of $H$-polarization the magnetic field is
polarized along the $z$-direction and the electric one confined to
the plane $XY$ (i.e., $H_{h,x} = H_{h,y} = E_{h,z} = 0$), with the
components of the latter being
\begin{equation}
 E_{h,x} = \frac{i}{\omega \epsilon_0}
  \frac{\partial H_{h,z}}{\partial y} , \qquad
 E_{h,y} = - \frac{i}{\omega \epsilon_0}
  \frac{\partial H_{h,z}}{\partial x} .
 \label{eq19}
\end{equation}
Substituting now these relations into the second line of (\ref{eq14})
yields
\begin{equation}
 \frac{\partial^2 H_{h,z}}{\partial x^2}
  + \frac{\partial^2 H_{h,z}}{\partial y^2} + k^2 H_{h,z} = 0 ,
 \label{eq20}
\end{equation}
which allows us to characterize $H$-polarization as
\begin{eqnarray}
 {\bf E}_h & = &  E_{h,x} \hat{\bf x} + E_{h,y} \hat{\bf y} ,
 \label{eq23} \\
 {\bf H}_h & = &  H_{h,z} \hat{\bf z} ,
 \label{eq24}
\end{eqnarray}
with $H_{h,z}$ satisfying the Helmholtz equation (\ref{eq20}).
Therefore, any general (time-independent) solution will be expressible
as
\begin{eqnarray}
 {\bf E} & = &  {\bf E}_e + {\bf E}_h
  = {\bf E}_e + \frac{i }{\omega \epsilon_0}
 \left[ \nabla \times {\bf H}_h \right],
 \label{eq25} \\
 {\bf H} & = & {\bf H}_e + {\bf H}_h
  = - \frac{i}{\omega \mu_0}
  \left[ \nabla \times {\bf E}_e \right] + {\bf H}_h .
 \label{eq26}
\end{eqnarray}

Since $E_z$ and $H_z$ satisfy Helmholtz's equation, consider now that
both are proportional to a scalar field, $\Psi ({\bf r})$, which also
satisfies this equation, i.e.,
\begin{eqnarray}
 {\bf E}_e & = & \alpha \Psi \hat{\bf z} ,
 \label{eq27} \\
 {\bf H}_h & = &\beta \sqrt{\frac{\epsilon_0}{\mu_0}} \
  e^{i\phi} \Psi \hat{\bf z} .
 \label{eq28}
\end{eqnarray}
Here, $\alpha$ and $\beta$ are real quantities and the phase shift
between both components is given by $\phi$; (\ref{eq6}) has been used
to obtain the correct dimensionality in the r.h.s.\ of (\ref{eq28}).
If (\ref{eq27}) and (\ref{eq28}) are substituted into Eqs.~(\ref{eq25})
and (\ref{eq26}), respectively, the latter become
\begin{eqnarray}
 {\bf E} & = &
  \frac{i \beta e^{i \phi}}{k}
    \frac{\partial \Psi}{\partial y} \ \hat{\bf x}
  - \frac{i \beta e^{i \phi}}{k}
    \frac{\partial \Psi}{\partial x} \ \hat{\bf y}
  + \alpha \Psi \hat{\bf z} ,
 \label{eq29} \\
 {\bf H} & = &
  - \frac{i \alpha}{\omega \mu_0}
    \frac{\partial \Psi}{\partial y} \ \hat{\bf x}
  + \frac{i \alpha}{\omega \mu_0}
    \frac{\partial \Psi}{\partial x} \ \hat{\bf y}
  + \frac{k \beta e^{i \phi}}{\omega \mu_0} \ \Psi \hat{\bf z} ,
 \label{eq30}
\end{eqnarray}
with their time-dependent counterparts being
\begin{eqnarray}
 \tilde{\bf E}({\bf r},t) & = &
  \left[  \frac{i \beta e^{i \phi}}{k}
    \frac{\partial \Psi}{\partial y} \ \hat{\bf x}
 - \frac{i \beta e^{i \phi}}{k}
    \frac{\partial \Psi}{\partial x} \ \hat{\bf y}
 + \alpha \Psi \hat{\bf z}
  \right] e^{-i \omega t} . \nonumber \\ & &
 \label{eq31}\\
 \tilde{\bf H}({\bf r},t) & = &
  \left[ - \frac{i \alpha}{\omega \mu_0}
    \frac{\partial \Psi}{\partial y} \ \hat{\bf x}
 + \frac{i \alpha}{\omega \mu_0}
    \frac{\partial \Psi}{\partial x} \ \hat{\bf y}
 + \frac{k \beta e^{i \phi}}{\omega \mu_0} \ \Psi \hat{\bf z}
  \right] e^{-i \omega t} , \nonumber \\ & &
 \label{eq32}
\end{eqnarray}
Equations (\ref{eq31}) and (\ref{eq32}) are general time-dependent
solutions for a problem which can be described in terms of
superpositions, as also happens with (\ref{eq25}) and (\ref{eq26}).
Once this set of equations is set up, the whole problem reduces to
finding $\Psi$ and its propagation along $x$ and $y$ (by the above
hypothesis, the set of equations does not depend on $z$), which is a
boundary condition problem.


\section{\label{sec3} Incident EM plane wave, its polarization and
photon trajectories}

Before reaching the two slits, we assume that both the electric and
magnetic fields propagate along the $y$-direction.
Moreover, we also assume that the incident scalar field is a
monochromatic plane wave,
\begin{equation}
 \Psi_0 ({\bf r}) = e^{i k y} .
 \label{eq33}
\end{equation}
Of course, this is just a simplification in order to develop some
analytical expressions, where we have assumed that the incident wave
is locally plane in a region that will cover the two slits.
However, in a more realistic case, one could consider a more general
incident wave function of a limited extension, like a beam or a wave
packet.
In the case of massive particles, the effects associated with incident
wave functions of limited extension in two-slit experiments have been
studied in \cite{sanz2a,sanz2b} (similarly, a study of the effects of
the outgoing diffracted waves can be found in \cite{sanz1b}).

Introducing (\ref{eq33}) into (\ref{eq31}) and (\ref{eq32}) yields
\begin{eqnarray}
 \tilde{\bf E}_0 ({\bf r},t) & = &
 \left[ E_{0,h,x} \hat{\bf x} + E_{0,e,z} \hat{\bf z} \right]
  e^{-i\omega t}
 \nonumber \\
 & = & \left[ - \beta e^{i(ky + \phi)} \hat{\bf x}
  + \alpha e^{iky} \hat{\bf z} \right] e^{-i\omega t} ,
 \label{eq34} \\
 \tilde{\bf H}_0 ({\bf r},t) & = &
  \left[ H_{0,h,x} \hat{\bf x} + H_{0,e,z} \hat{\bf z} \right]
   e^{-i\omega t}
  \nonumber \\
 & = & \sqrt {\frac {\epsilon_0}{ \mu_0}}
 \left[ \alpha e^{iky} \hat{\bf x}
  + \beta e^{i(ky + \phi)} \hat{\bf z} \right] e^{-i\omega t} .
 \label{eq35}
\end{eqnarray}
From these solutions, it follows that the polarization is going to play
an important role in the interference patterns observed, and also in
the topology displayed by the photon trajectories.

Consider the electric field (\ref{eq34}), whose real components are
\begin{eqnarray}
 \tilde{E}_{0,h,x}^r & = & -\beta \cos (k y - \omega t + \phi) ,
 \label{eq36} \\
 \tilde{E}_{0,e,z}^r & = & \alpha \cos (k y - \omega t) .
 \label{eq37}
\end{eqnarray}
Since the magnetic field displays the same polarization properties
as the electric field due to their relationship through the Maxwell
equations (\ref{eq5}) and (\ref{eq6}), we will only argue in terms
of the electric field without loss of generality.
Thus, expressing (\ref{eq36}) and (\ref{eq37}) as
\begin{eqnarray}
 \frac{\tilde{E}_{0,h,x}^r}{\beta}
   + \frac{\tilde{E}_{0,e,z}^r}{\alpha} \cos \phi & = &
    \sin (k y - \omega t) \sin \phi ,
 \label{eq38} \\
 \frac{\tilde{E}_{0,e,z}^r}{\alpha} \sin \phi & = &
   \cos (k y - \omega t) \sin \phi ,
 \label{eq39}
\end{eqnarray}
and then squaring and rearranging terms, we reach
\begin{equation}
 \left( \frac{\tilde{E}_{0,h,x}^r}{\beta} \right)^2
   + \left( \frac{\tilde{E}_{0,e,z}^r}{\alpha} \right)^2
   + 2 \left( \frac{\tilde{E}_{0,h,x}^r}{\beta} \right) \! \!
     \left( \frac{\tilde{E}_{0,e,z}^r}{\alpha} \right)
        \cos \phi = \sin^2 \phi .
 \label{eq40}
\end{equation}
According to this relation, several cases are possible depending on
the value of the phase-shift $\phi$:
\begin{itemize}
\item[(a)] When $\phi = 0$ or $\pi$,
\begin{equation}
 \left( \frac{\tilde{E}_{0,h,x}^r}{\beta} \pm
   \frac{\tilde{E}_{0,e,z}^r}{\alpha} \right)^2 = 0 \quad
  \Rightarrow \quad \frac{\tilde{E}_{0,h,x}^r}{\beta}
     = \mp \frac{\tilde{E}_{0,e,z}^r}{\alpha} .
 \label{eq41}
\end{equation}
This case describes {\it linear polarization}, for any arbitrary
$\alpha$ and $\beta$.

\begin{figure}
 \begin{center}
 \includegraphics[height=8cm,angle=-90]{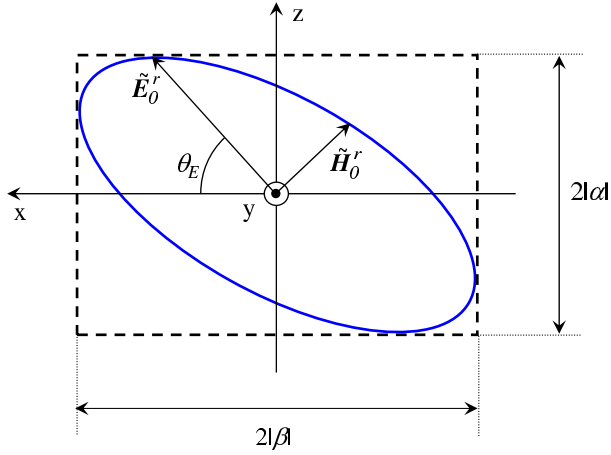}
 \caption{\label{fig0}
  Diagram showing the general elliptic motion described by the electric
  field due to its polarization (the same applies to the magnetic field,
  which is perpendicular).
  If $\phi = \pm \pi/2$ and $\alpha = \beta$, the ellipse becomes
  a circle (circular polarization), and if $\phi = 0$ or $\pi$, it
  reduces to a segment (linear polarization) regardless of the
  $\alpha$ and $\beta$ values.}
 \end{center}
\end{figure}

\item[(b)] For any other value of $\phi$, there is {\it elliptic
polarization}.
In this case, (\ref{eq40}) is the equation of an ellipse inscribed
in a rectangle parallel to the $XZ$ plane, with sides $2|\alpha|$
and $2|\beta|$ (see Fig.~\ref{fig0}).
The electric field (and also the magnetic one, which is perpendicular)
can move clockwise or anticlockwise, as seen by an observer toward whom
the EM wave is moving.
These rotations define, respectively, right-handed or left-handed
polarization states.
The polarization handedness can be determined by computing the time
derivative of the angle formed by the two components of the electric
field, $\theta_E=(\tan)^{-1}(\tilde{E}_{0,e,z}^r/\tilde{E}_{0,h,x}^r)$,
\begin{eqnarray}
 \frac{d\theta_E}{dt}
  & = & \frac{1}{1 + (\tilde{E}_{0,e,z}^r/\tilde{E}_{0,h,x}^r)^2} \
   \frac{d}{dt} \left( \frac{\tilde{E}_{0,e,z}^r}{\tilde{E}_{0,h,x}^r}
    \right) \nonumber \\
  & = & - \frac{\beta \sin \phi}{\alpha} \
   \frac{\omega \left[ 1 + \tan^2 (ky - \omega t ) \right]}
    {1 + (\tilde{E}_{0,e,z}^r/\tilde{E}_{0,h,x}^r)^2} .
\end{eqnarray}
As can be noticed, the information about the handedness is contained
in the prefactor of this expression, since the second factor (the
time-dependent one) is always positive.
Thus, if $\alpha$ and $\beta$ are chosen positive, $d\theta_E/dt$ will
be positive for $-\pi<\phi<0$ and negative for $0<\phi<\pi$.
In the first case, the field is right-handed polarized ($\theta_E$
increases with time) and, in the second one, it is left-handed
($\theta_E$ decreases with time).

\item[(c)] In the particular case $\phi = \pm \pi/2$, we have
\begin{equation}
 \left( \frac{\tilde{E}_{0,h,x}^r}{\beta} \right)^2 +
  \left( \frac{\tilde{E}_{0,e,z}^r}{\alpha} \right)^2 = 1 .
 \label{eq42}
\end{equation}
If $\alpha = \beta$, the ellipse described by (\ref{eq42}) reduces to
the equation of circle.
We then have {\it circular polarization}.
\end{itemize}

The particular values of $\phi$, $\alpha$ and $\beta$, which define
the polarization state of the incident plane wave, are very important
regarding the observation of interference patterns behind the slits
according to the Arago-Fresnel laws \cite{barakat}.
But they are also going to be very important with respect to the
topology of the corresponding photon trajectories, as shown below.

In Sec.~\ref{sec4}, we will tackle the calculation of the EM field and
photon trajectories behind gratings with $N$ and two slits.
In the second case, we will not only consider free passage through the
slits, but also when they are covered by linear polarizers parallel to
the $XZ$ plane, whose polarization axes are oriented along the $z$-axis
in one of the slits and along the $x$-axis in the other.
Moreover, and without loss of generality, we are going to assume that
the polarizer oriented along the $z$-axis produces $E$-polarization,
while the polarizer along the $x$-axis gives rise to $H$-polarization.
As can be shown, the action of these polarizers on an incident plane
wave which propagates along the $y$-axis, such as (\ref{eq33}),
justifies this specific labelling.
Note that, after passing through such polarizers, the EM field
described by Eqs.~(\ref{eq34}) and (\ref{eq35}) gives rise to
transmitted fields, ${\bf E}_{tr}$ and ${\bf H}_{tr}$, with the
following polarizations:
\begin{itemize}
\item[a)] If the polarization axis is oriented along the $x$-axis:
\begin{eqnarray}
 {\bf E}_{tr}({\bf r}) & = & - \beta e^{i(ky + \phi)} \hat{\bf x}
  = E_{0,h,x} \hat{\bf x} ,
 \label{eqpx1}\\
 {\bf H}_{tr}({\bf r}) & = & \sqrt{\frac{\epsilon_0}{\mu_0}} \
  \beta e^{i(ky + \phi)} \hat{\bf z} = H_{0,h,z} \hat{\bf z} .
 \label{eqpx2}
\end{eqnarray}

\item[b)] If the polarization axis is oriented along the $z$-axis:
\begin{eqnarray}
 {\bf E}_{tr}({\bf r}) & = & \alpha e^{iky} \hat{\bf z}
  = E_{0,e,z} \hat{\bf z} ,
 \label{eqpz1} \\
 {\bf H}_{tr}({\bf r}) & = & \sqrt{\frac{\epsilon_0}{\mu_0}} \
  \alpha e^{iky} \hat{\bf x} = H_{0,e,x} \hat{\bf x} .
 \label{eqpz2}
\end{eqnarray}
\end{itemize}

Regarding the photon trajectories, before the EM field reaches the
grating, they are obtained from
\begin{eqnarray}
 {\bf S}_0({\bf r}) & = & \frac{1}{2} \
  {\rm Re} [ {\bf E}_0({\bf r}) \times {\bf H}_0^*({\bf r}) ]
  \nonumber \\
  & = & \frac{1}{2} \sqrt{\frac{\epsilon_0}{\mu_0}} \
    (\alpha^2 + \beta^2) \hat{\bf y} ,
 \label{eq43} \\
 U_0({\bf r}) & = & \frac{1}{4} \left[
  \epsilon_0 {\bf E}({\bf r}) \cdot {\bf E}^*({\bf r})
  + \mu_0 {\bf H}({\bf r}) \cdot {\bf H}^*({\bf r}) \right]
  \nonumber \\
  & = & \frac{\epsilon_0}{2} \ (\alpha^2 + \beta^2) ,
 \label{eq43b}
\end{eqnarray}
where the time-independent parts of (\ref{eq34}) and (\ref{eq35}) have
been considered.
Substituting (\ref{eq43}) and (\ref{eq43b}) into (\ref{eq44}), we
obtain
\begin{equation}
 \frac{d{\bf r}}{ds} = \hat{\bf y} ,
 \label{eq46}
\end{equation}
which, after integration, yields
\begin{eqnarray}
 x(s) & = & x_0 , \quad z(s) = z_0 ,
 \label{eq49} \\
 y(s) & = & y_0 + s .
 \label{eq50}
\end{eqnarray}
That is, the EME flow evolves along the $y$-direction and, therefore,
photons pursue straight lines parallel to the $y$-axis.
Since we are in vacuum, we can assume that the distance $s$ travelled
by a photon during a time $t$ is given by $s = c t$ and, therefore,
(\ref{eq50}) can also be expressed as
\begin{equation}
 y(t) = y_0 + c t .
 \label{eq52}
\end{equation}


\section{\label{sec4} EM field behind a grating}

According to Sec.~\ref{sec2}, the EM field behind a grating can be
described by Eqs.~(\ref{eq29}) and (\ref{eq30}) (or Eqs.~(\ref{eq31})
and (\ref{eq32}), respectively, if we consider time-dependence), with
the scalar function $\Psi ({\bf r}) = \Psi(x,y)$ satisfying both
Helmholtz's equation and the boundary conditions at the grating.
If the incident EM field is monochromatic, we can also assume that the
incident scalar function is given by (\ref{eq33}).

Traditionally, the exact solution for the slit array that we are going
to consider here arises from Helmholtz's equation (see Eqs.~(\ref{eq7})
and (\ref{eq8})) and is expressed as a Fresnel-Kirchhoff integral
\cite{bornwolf},
\be
 \Psi(x,y) = \sqrt{\frac{k}{2\pi y}} \ e^{-i\pi/4} e^{iky} \! \!
  \int_{-\infty}^\infty \Psi(x',0^+) e^{ik(x - x')^2/2y} dx' ,
 \label{helm0}
\ee
where $\Psi(x',0^+)$ denotes the wave function right behind the two
slits.
Then, after assuming some appropriate approximations, expressions valid
to describe Fresnel and Fraunhofer diffraction can be obtained from the
corresponding general solutions.
Alternatively, Arsenovi\'c {\it et al.}\ \cite{mirjana3} have shown
that the solution behind a grating can also be expressed as a
superposition of transverse modes of the fields multiplied by an
exponential function of the longitudinal coordinate, i.e.,
\ba
 \Psi(x,y) & = & \frac{1}{\sqrt{2\pi}} \ \! e^{iky} \! \!
  \int c(k_x) \ \!  e^{ik_x x - ik_x^2 y/2k} dk_x ,
 \label{helm1}
\ea
where
\be
 c(k_x) = \frac{1}{\sqrt{2\pi}} \ \! \int
   \Psi(x,0^+) \ \! e^{-ik_x x} dx .
 \label{helm2}
\ee
As also shown by Arsenovi\'c {\it et al.}~\cite{mirjana4,mirjana5},
the solution (\ref{helm1}) is equivalent to the Fresnel-Kirchhoff
integral (\ref{helm0}) whenever the wave number associated with the
transverse mode $k_x$ of a general slit satisfies the condition
$k \gg k_x$, which occurs in most cases of physical interest.

For a grating which is totally transparent inside the slits and
completely absorbing outside them, the boundary conditions are:
$\Psi(x,0^+) = 0$ for any $x$ belonging to the slit support and
$\Psi(x,0^+) = \Psi(x,0^-)$ for any $x$ within the apertures, with
$\Psi(x,0^-)$ being the wave function incident on the grating, here
given by (\ref{eq33}).
Thus, in the case of an incident beam illuminating $N$ openings of
width $\delta$ and mutual distance $d$, a simple analytical calculation
\cite{mirjana6} renders
\be
 c_N(k_x) = \sqrt{\frac{\delta}{2\pi N}}
  \left[ \frac{\sin (k_x\delta/2)}{k_x\delta/2} \right]
  \left[ \frac{\sin (N k_x d/2)}{\sin (k_x d/2)} \right] .
 \label{helm5}
\ee
In the particular case $N = 2$, i.e., the well-known double-slit
experiment, it is useful to express Eqs.~(\ref{helm0}) and
(\ref{helm2}) as
\begin{eqnarray}
 \Psi(x,y) & = & \sqrt{\frac{k}{2\pi y}} \ e^{-i\pi /4}e^{iky}
   \int_{A_1} \Psi(x',0^-) e^{ik(x - x')^2/2y} dx'
 \nonumber\\
  & & + \sqrt{\frac{k}{2\pi y}} \ e^{-i\pi /4}e^{iky}
   \int_{A_2} \Psi(x',0^-) e^{ik(x - x')^2/2y} dx'
 \nonumber\\
  & \equiv & \psi_1(x,y) + \psi_2(x,y) ,
 \label{eqsu1}
\end{eqnarray}
and
\begin{eqnarray}
 c_2(k_x) & = & \frac{1}{\sqrt{2\pi}}
   \int_{A_1}\Psi(x',0^-)e^{-i k_x x}dx  \nonumber\\
 & & + \frac{1}{\sqrt{2\pi}} \int_{A_2}\Psi(x',0^-) e^{-i k_x x}dx
  \nonumber\\
 & \equiv & \frac{1}{\sqrt{2}} \left( c_{1,d/2} + c_{1,-d/2} \right) ,
 \label{eqsu2}
\end{eqnarray}
respectively, where $\psi_1$ refers to the scalar field coming from
slit 1, centered at $x = d/2$, and $\psi_2$ is the scalar field coming
from slit 2, at $x = - d/2$.
After carrying out each integral in (\ref{eqsu2}), we obtain
\begin{eqnarray}
 c_{1,d/2}(k_x) & = & \sqrt{\frac{2}{\pi\delta}}
  \left[ \frac{\sin (k_x\delta/2)}{k_x} \right]
   e^ {-i k_x d/2} ,
 \nonumber\\
 c_{1,-d/2}(k_x) & = & \sqrt{\frac{2}{\pi\delta}}
  \left[ \frac{\sin (k_x\delta/2)}{k_x} \right]
   e^ {i k_x d/2} ,
 \label{eqc1}
\end{eqnarray}
which, when they are added, yield
\be
 c_2(k_x) = \sqrt{\frac{\delta}{\pi}}
  \left[ \frac{\sin (k_x\delta/2)}{k_x\delta/2} \right]
   \cos (k_x d/2) .
 \label{eqc2}
\ee

In the space behind the grating, the EM field is given by
Eqs.~(\ref{eq29}) and (\ref{eq30}), where $\Psi$ is described by
(\ref{eqsu1}).
That is, the resulting EM field behind the grating consists of a
superposition of two fields propagating from each slit, which
reads as
\begin{eqnarray}
  {\bf E} & = &
  \frac{i \beta e^{i \phi}}{k}
    \frac{\partial \psi_1}{\partial y} \ \hat{\bf x}
 - \frac{i \beta e^{i \phi}}{k}
    \frac{\partial \psi_1}{\partial x} \ \hat{\bf y}
 + \alpha \psi_1 \hat{\bf z} \nonumber \\
 & & + \frac{i \beta e^{i \phi}}{k}
    \frac{\partial \psi_2}{\partial y} \ \hat{\bf x}
 - \frac{i \beta e^{i \phi}}{k}
    \frac{\partial \psi_2}{\partial x} \ \hat{\bf y}
 + \alpha \psi_2 \hat{\bf z} \nonumber \\
 & \equiv & {\bf E}_1 + {\bf E}_2 ,
 \label{eqn65}\\
 {\bf H} & = &
  - \frac{i \alpha}{\omega \mu_0}
    \frac{\partial \psi_1}{\partial y} \ \hat{\bf x}
 + \frac{i \alpha}{\omega \mu_0}
    \frac{\partial \psi_1}{\partial x} \ \hat{\bf y}
 + \frac{k \beta e^{i \phi}}{\omega \mu_0} \ \psi_1 \hat{\bf z}
 \nonumber \\
 & & - \frac{i \alpha}{\omega \mu_0}
    \frac{\partial \psi_2}{\partial y} \ \hat{\bf x}
 + \frac{i \alpha}{\omega \mu_0}
    \frac{\partial \psi_2}{\partial x} \ \hat{\bf y}
 + \frac{k \beta e^{i \phi}}{\omega \mu_0} \ \psi_2 \hat{\bf z}
 \nonumber \\
 & \equiv & {\bf H}_1 + {\bf H}_2 .
 \label{eqn66}
\end{eqnarray}


\section{\label{sec5} Photon trajectories behind the two slits}

In order to obtain the photon trajectories behind the grating, first we
substitute (\ref{eq29}) and (\ref{eq30}) into (\ref{eq9}), which yields
the components of the Poynting vector along the different directions,
\begin{eqnarray}
 S_x & = & \frac{i(\alpha^2 + \beta^2)}{4\omega\mu_0}
  \left( \Psi \frac{\partial \Psi^*}{\partial x}
  - \Psi^* \frac{\partial \Psi}{\partial x} \right) ,
 \label{eq54} \\
 S_y & = & \frac{i(\alpha^2 + \beta^2)}{4\omega\mu_0}
  \left( \Psi \frac{\partial \Psi^*}{\partial y}
  - \Psi^* \frac{\partial \Psi}{\partial y} \right) ,
 \label{eq55} \\
 S_z & = & - \frac{i \alpha \beta \sin \phi}{2k\omega\mu_0}
  \left(
    \frac{\partial \Psi}{\partial x} \frac{\partial \Psi^*}{\partial y}
  - \frac{\partial \Psi}{\partial y} \frac{\partial \Psi^*}{\partial x}
   \right) .
 \label{eq56}
\end{eqnarray}
Proceeding similarly with (\ref{eq45}) leads us to the EME density,
\begin{equation}
 U = \frac{(\alpha^2 + \beta^2)}{4\omega^2\mu_0} \left(
   \frac{\partial \Psi}{\partial x} \frac{\partial \Psi^*}{\partial x}
 + \frac{\partial \Psi}{\partial y} \frac{\partial \Psi^*}{\partial y}
 + k^2 \Psi \Psi^* \right) ,
 \label{eq57}
\end{equation}
which describes the interference pattern at the observation screen.

Before computing the photon trajectories, it is interesting to note the
following feature about the interference pattern.
Consider (\ref{eq57}) expressed in terms of the two scalar fields,
$\psi_1$ and $\psi_2$, i.e.,
\begin{eqnarray}
 U & = & \frac{(\alpha^2 + \beta^2)}{4\omega^2\mu_0} \left(
   \frac{\partial \psi_1}{\partial x}
     \frac{\partial \psi_1^*}{\partial x}
 + \frac{\partial \psi_1}{\partial y}
     \frac{\partial \psi_1^*}{\partial y}
 + k^2 \psi_1 \psi_1^* \right)
 \nonumber \\
 & & + \frac{(\alpha^2 + \beta^2)}{4\omega^2\mu_0} \left(
   \frac{\partial \psi_2}{\partial x}
     \frac{\partial \psi_2^*}{\partial x}
 + \frac{\partial \psi_2}{\partial y}
     \frac{\partial \psi_2^*}{\partial y}
 + k^2 \psi_2 \psi_2^* \right)
 \nonumber \\
 & & + \frac{(\alpha^2 + \beta^2)}{4\omega^2\mu_0} \left(
   \frac{\partial \psi_1}{\partial x}
     \frac{\partial \psi_2^*}{\partial x}
 + \frac{\partial \psi_1}{\partial y}
     \frac{\partial \psi_2^*}{\partial y}
 + k^2 \psi_1 \psi_2^* \right)
 \nonumber \\
 & & + \frac{(\alpha^2 + \beta^2)}{4\omega^2\mu_0} \left(
   \frac{\partial \psi_2}{\partial x}
     \frac{\partial \psi_1^*}{\partial x}
 + \frac{\partial \psi_2}{\partial y}
     \frac{\partial \psi_1^*}{\partial y}
 + k^2 \psi_2 \psi_1^* \right) .
 \label{eq67}
\end{eqnarray}
In short-hand notation, (\ref{eq67}) can also be expressed as
\begin{equation}
 U = U_1 + U_2 + U_{12} ,
 \label{eq68}
\end{equation}
where $U_1$ and $U_2$ are the EME densities associated with $\psi_1$
and $\psi_2$ (the first and second terms in (\ref{eq67})),
respectively.
On the other hand, $U_{12}$ (the last two terms in (\ref{eq67})) is the
EME density arising form the {\it interference} of these waves.
Since the polarization part (prefactor in terms of $\alpha$ and
$\beta$) and the space part (depending on $\Psi$) appear factorized in
(\ref{eq57}), the interference pattern observed will not depend on
the polarization state of the incident field, in agreement with the
Arago-Fresnel laws.

Substituting now (\ref{eq54})-(\ref{eq57}) into (\ref{eq44}) renders
the corresponding trajectory equations along each direction,
\be
 \frac{dx}{ds} = ik \left\{
  \frac{\displaystyle \Psi \frac{\partial \Psi^*}{\partial x}
  - \Psi^* \frac{\partial \Psi}{\partial x}}
  {\displaystyle
   \frac{\partial \Psi}{\partial x} \frac{\partial \Psi^*}{\partial x}
 + \frac{\partial \Psi}{\partial y} \frac{\partial \Psi^*}{\partial y}
 + k^2 \Psi \Psi^*} \right\} ,
 \label{eq54b}
\ee
\be
 \frac{dy}{ds} = ik \left\{
  \frac{\displaystyle \Psi \frac{\partial \Psi^*}{\partial y}
  - \Psi^* \frac{\partial \Psi}{\partial y}}
  {\displaystyle
   \frac{\partial \Psi}{\partial x} \frac{\partial \Psi^*}{\partial x}
 + \frac{\partial \Psi}{\partial y} \frac{\partial \Psi^*}{\partial y}
 + k^2 \Psi \Psi^*} \right\} ,
 \label{eq55b}
\ee
\be
 \frac{dz}{ds} = - \frac{2i\alpha\beta\sin\phi}{(\alpha^2 + \beta^2)}
  \left\{ \frac{\displaystyle
   \frac{\partial \Psi}{\partial x} \frac{\partial \Psi^*}{\partial y}
 - \frac{\partial \Psi}{\partial y} \frac{\partial \Psi^*}{\partial x}}
  {\displaystyle
   \frac{\partial \Psi}{\partial x} \frac{\partial \Psi^*}{\partial x}
 + \frac{\partial \Psi}{\partial y} \frac{\partial \Psi^*}{\partial y}
 + k^2 \Psi \Psi^*} \right\} .
 \label{eq56b}
\ee
As can be noticed from these expressions, all the information about the
polarization state of the diffracted EM wave is contained in the
prefactor of (\ref{eq56b}).
Thus, regardless of the polarization of the initial EM wave, since the
diffracted waves arising from each slit have the same polarization
state, one will always observe interference fringes, which is in
agreement with the Arago-Fresnel laws \cite{barakat}.

\begin{figure}
 \begin{center}
 \includegraphics[width=7.5cm]{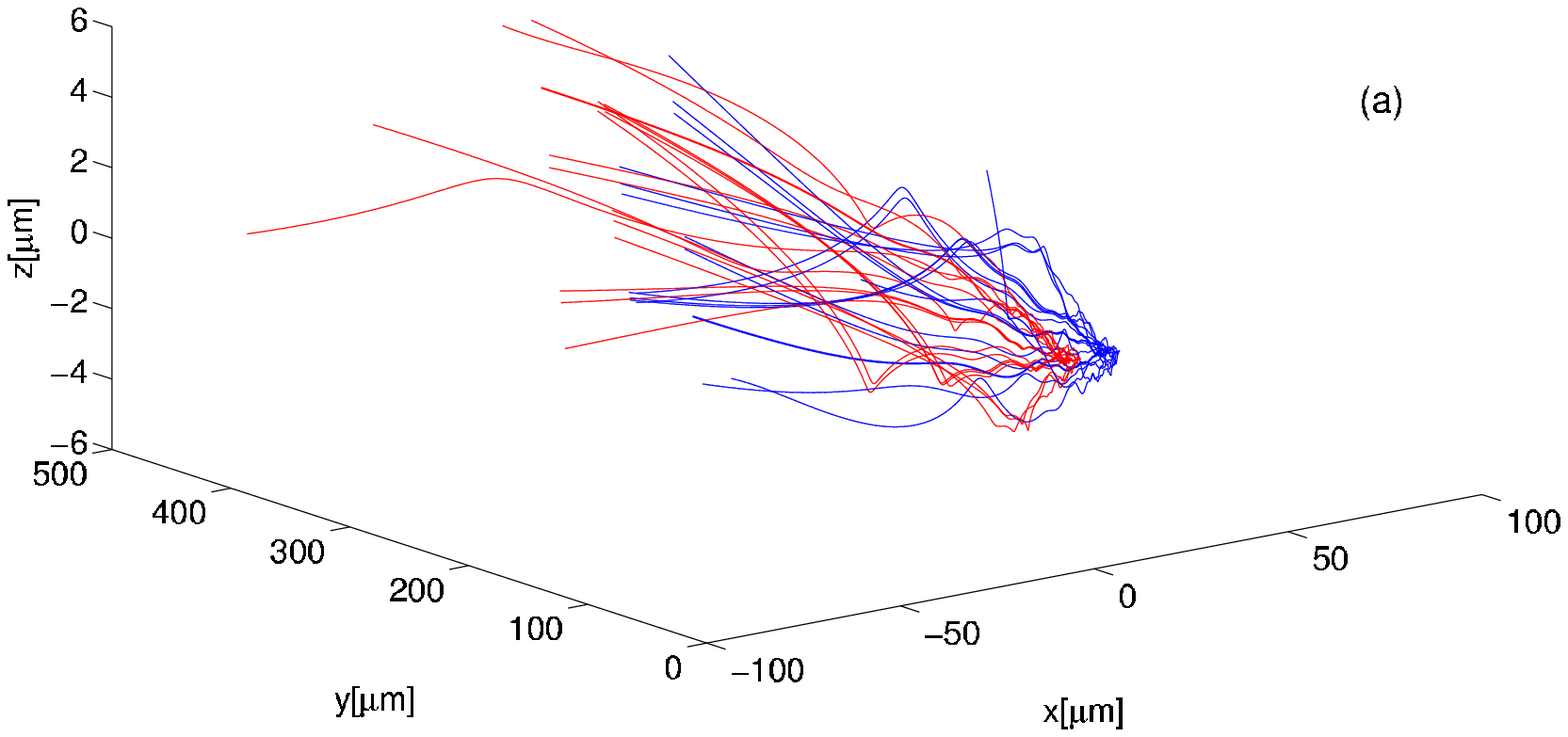}
 \includegraphics[width=7.5cm]{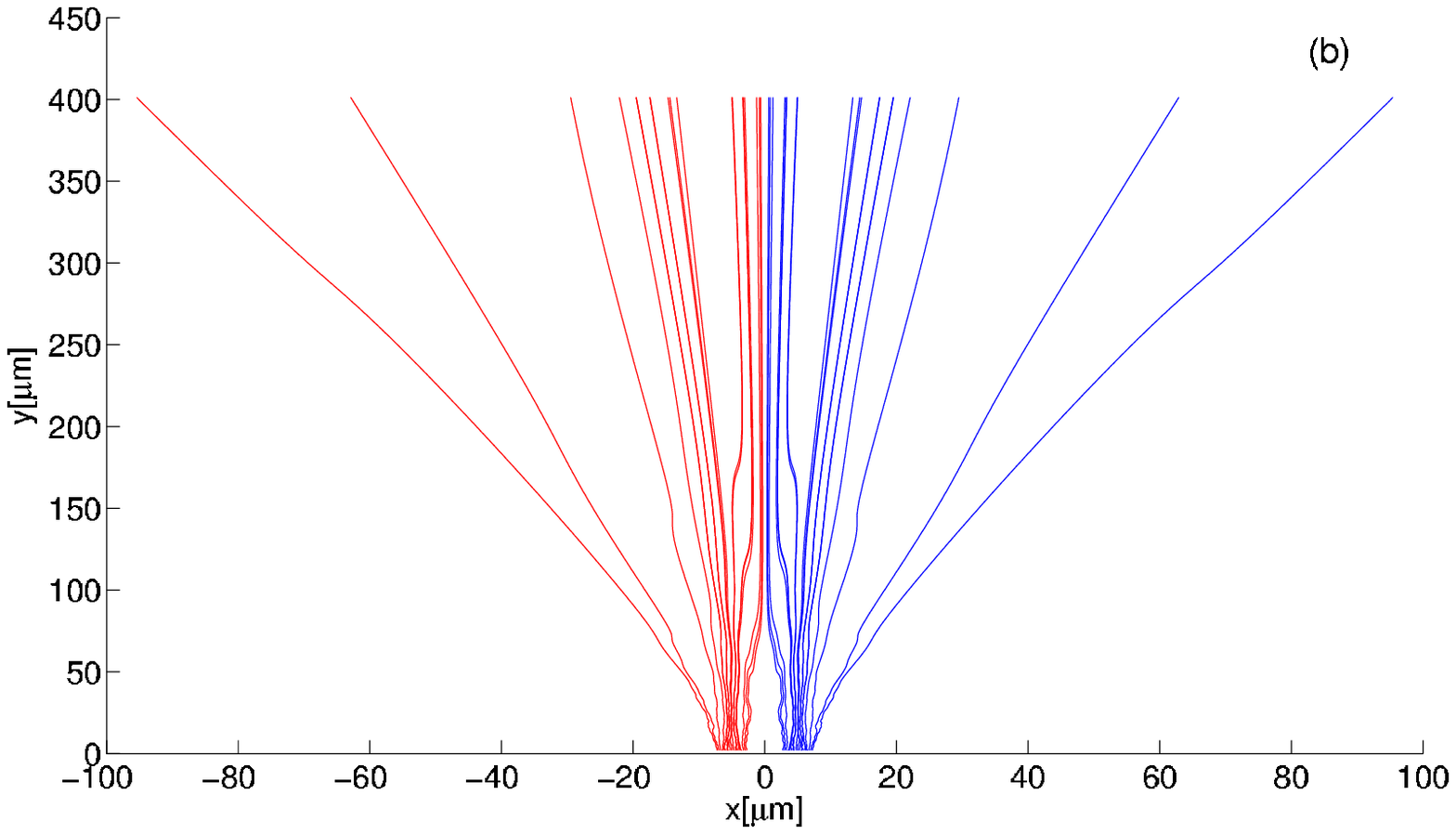}
 \includegraphics[width=7.5cm]{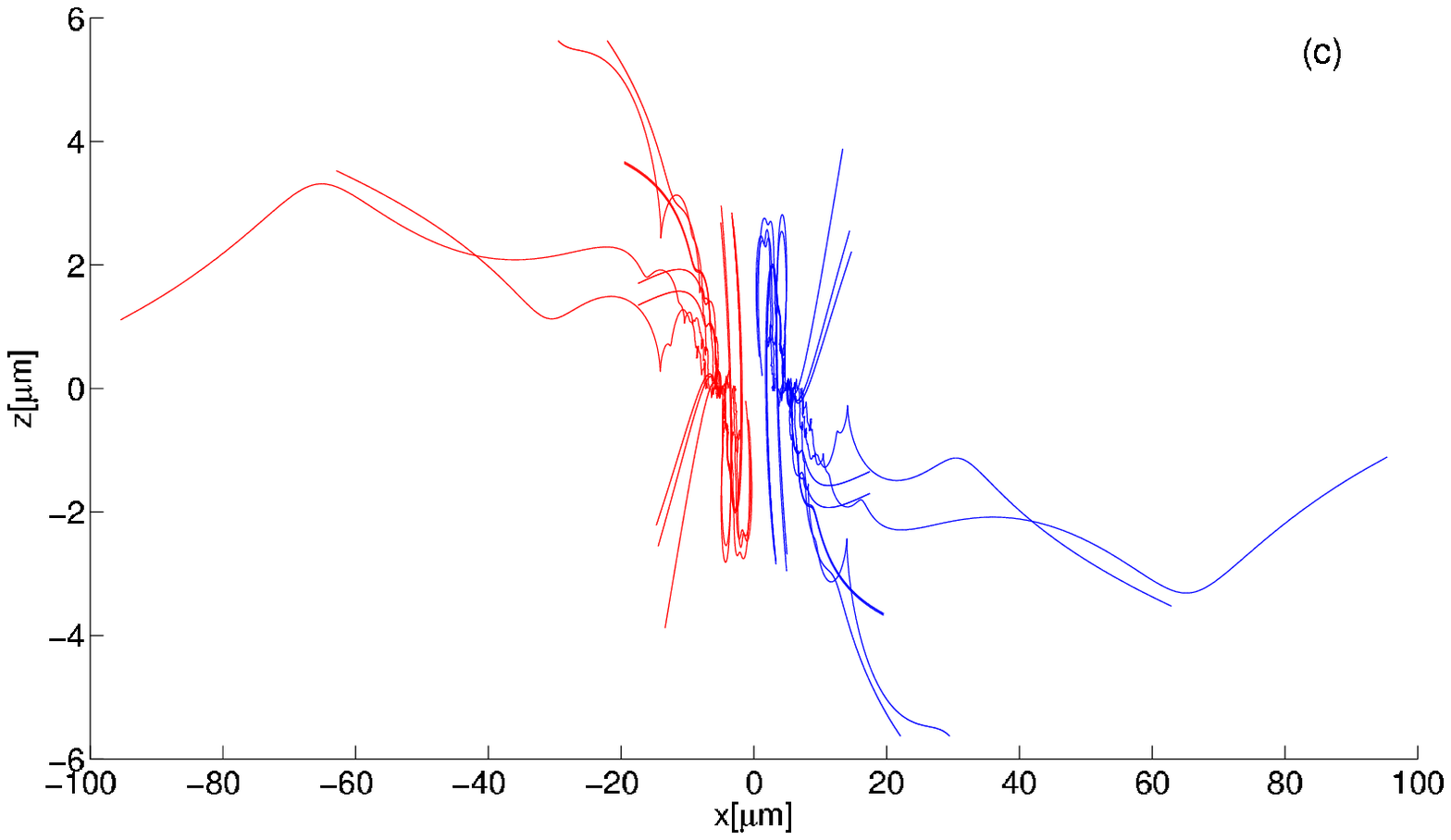}
 \includegraphics[width=7.5cm]{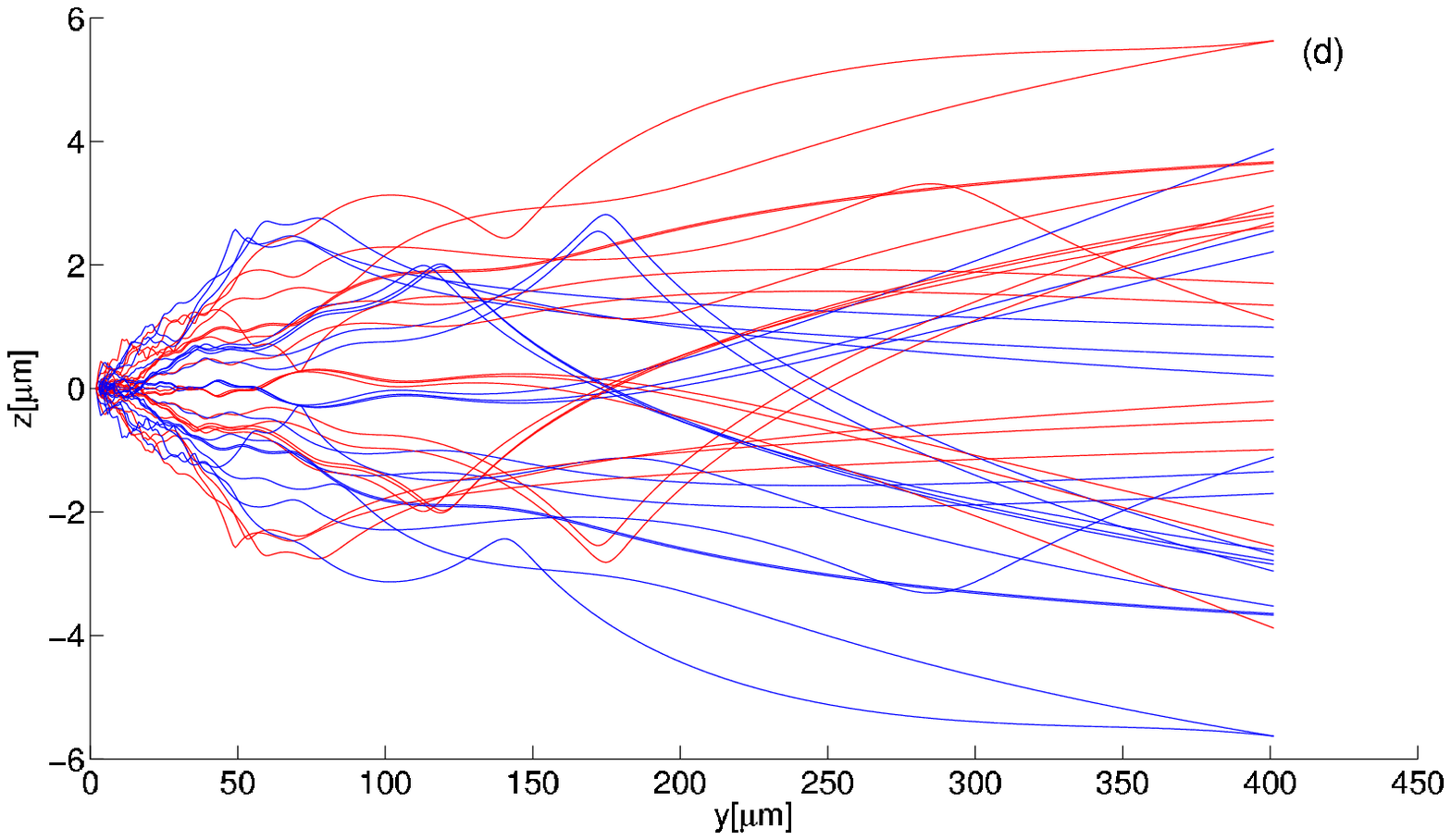}
 \caption{\label{fig1}
  Photon trajectories (15 for each slit) behind a two-slit grating
  associated with an incident EM plane wave is circularly polarized
  ($\alpha = \beta$, $\phi=\pi/2$): (a) 3D view, (b) $XY$ projection,
  (c) $XZ$ projection and (d) $YZ$ projection.
  The parameters considered in the simulation are: $\lambda = 500$~nm,
  $d = 20 \lambda = 10$~$\mu$m and $\delta = d/2 = 5$~$\mu$m.}
 \end{center}
\end{figure}

In the case of linear polarization, (\ref{eq56b}) vanishes
\cite{asanz-phys-scr} and we can solve the photon-trajectory equations
simply by parametrizing, for example, $y$ as a function of $x$, i.e.,
\begin{equation}
 \frac{dy}{dx} = \frac{\displaystyle
   \left( \Psi \frac{\partial \Psi^*}{\partial y}
  - \Psi^* \frac{\partial \Psi}{\partial y} \right)}
  {\displaystyle \left( \Psi \frac{\partial \Psi^*}{\partial x}
    - \Psi^* \frac{\partial \Psi}{\partial x} \right)} ,
 \label{eq58}
\end{equation}
while the solution of (\ref{eq56}) is simply $z = z_0$.
On the contrary, in the case of elliptic polarization, the
$z$-component does play an important role, as can be noticed
when the photon-trajectory equations are computed,
\begin{eqnarray}
 \frac{dy}{dx} & = & \frac{\displaystyle
   \left( \Psi \frac{\partial \Psi^*}{\partial y}
  - \Psi^* \frac{\partial \Psi}{\partial y} \right)}
  {\displaystyle \left( \Psi \frac{\partial \Psi^*}{\partial x}
    - \Psi^* \frac{\partial \Psi}{\partial x} \right)} ,
 \label{eq59} \\
 \frac{dz}{dx} & = &
 - \frac{2 \alpha \beta \sin \phi}{(\alpha^2 + \beta^2) k}
  \left\{ \frac{\displaystyle
    \frac{\partial \Psi}{\partial x} \frac{\partial \Psi^*}{\partial y}
  - \frac{\partial \Psi}{\partial y} \frac{\partial \Psi^*}{\partial x}
   }
  {\displaystyle \Psi \frac{\partial \Psi^*}{\partial x}
    - \Psi^* \frac{\partial \Psi}{\partial x} } \right\} .
 \label{eq60}
\end{eqnarray}
In Fig.~\ref{fig1}, the photon trajectories associated with the
diffraction of an incident EM field circularly polarized ($\alpha =
\beta$, $\phi=\pi/2$) are plotted.
The parameters considered in the simulation are: $\lambda = 500$~nm,
$d = 20 \lambda = 10$~$\mu$m and $\delta = d/2 = 5$~$\mu$m.
The projections of these flow lines on the $XY$ plane, shown in
Fig.~\ref{fig1}(b), are identical to those for incident linearly
polarized light \cite{asanz-phys-scr} (note that (\ref{eq59}) is
exactly the same as (\ref{eq58})).
As shown elsewhere \cite{asanz-jpa} within the context of Bohmian
mechanics, for equations like (\ref{eq59}), which describes the
photon-trajectory projections on the $XY$, trajectories exiting through
one slit never cross the trajectories coming up from the other one.
Moreover, if both slits are identical (here this means they have the
same width and transmission function), the EME fluxes coming out from
each slit are symmetric with respect to the axis $y=0$.
Now, although neither the electric nor the magnetic field depend on
the $z$-coordinate, the photon trajectories display some remarkable
features along this direction, as seen in Figs.~\ref{fig1}(c) and
\ref{fig1}(d).
This is an effect of having circular (or elliptical, in general)
polarization, which vanishes in the case of linear polarization,
when $\phi = 0$ or $\pi$.

\begin{figure}
 \begin{center}
 \includegraphics[width=7.5cm]{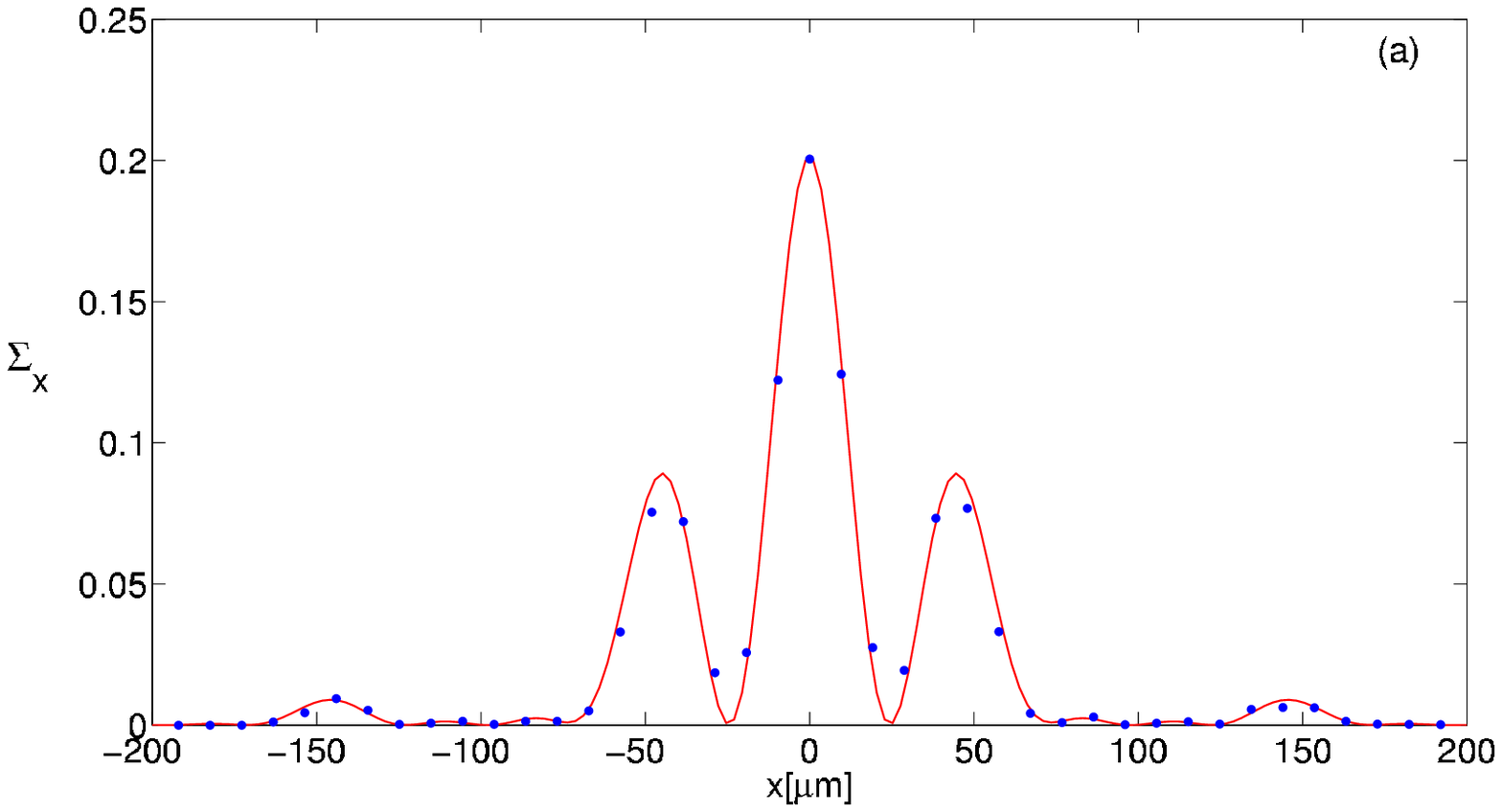}
 \includegraphics[width=7.5cm]{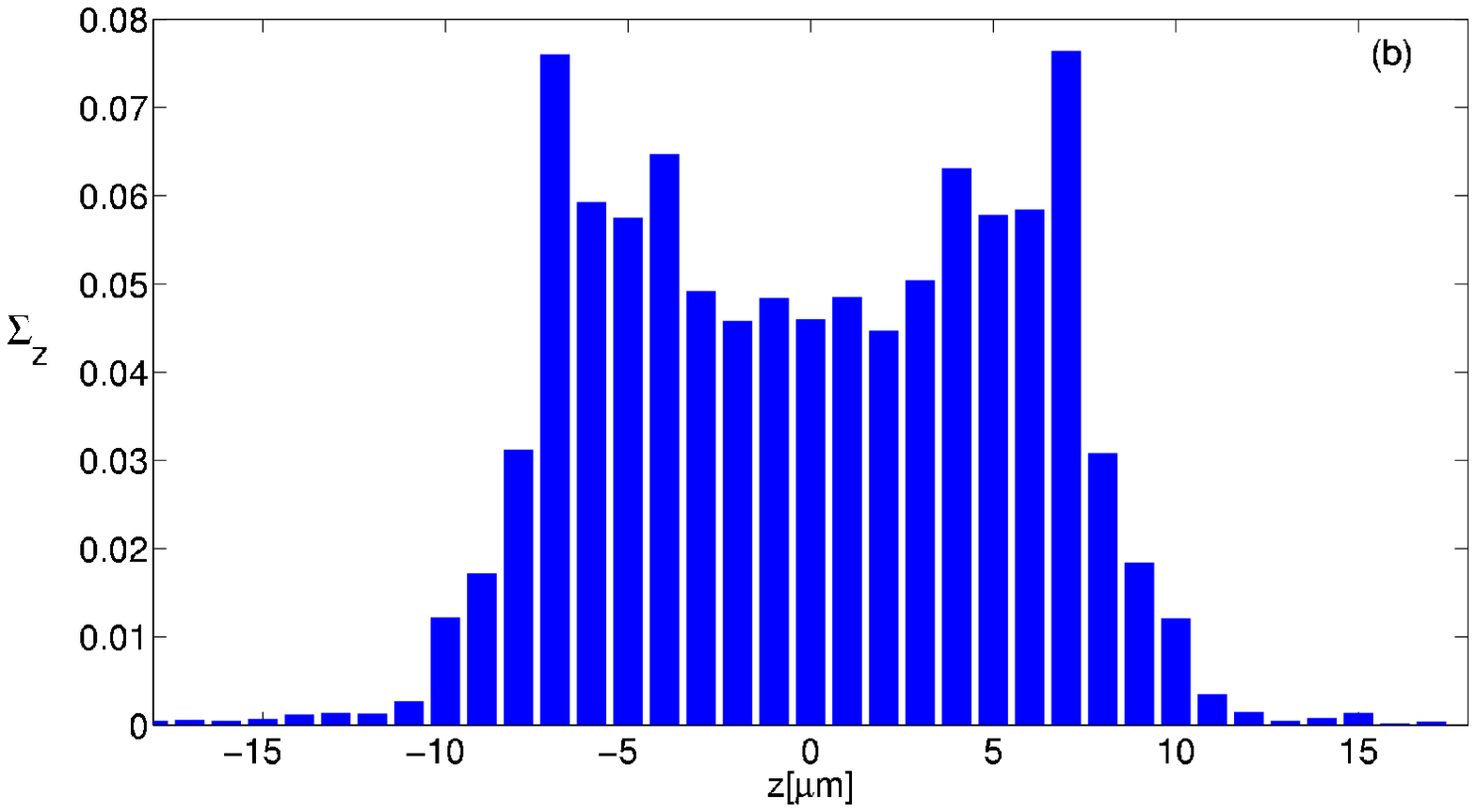}
 \caption{\label{fig2}
  Histograms built up by counting the end points of individual photon
  trajectories associated with the incident EM field circularly
  polarized ($\alpha = \beta$, $\phi=\pi/2$) of Fig.~\ref{fig1}.
  The detection screen is at $L = 1$~mm from a two-slit grating and the
  histograms represent counts: (a) along the $x$-direction and (b)
  along the $z$-direction.
  In part (a), the red solid line indicates the theoretical curve
  predicted by standard Electromagnetism, according to (\ref{eq57}).
  The parameters considered in the simulation are: $\lambda = 500$~nm,
  $d = 20 \lambda = 10$~$\mu$m and $\delta = d/2 = 5$~$\mu$m.
  The total number of trajectories considered is 5,000, with initial
  conditions homogenously distributed along each slit.}
 \end{center}
\end{figure}

In order to understand the somewhat unexpected motion along the
$z$-direction, let us go back to (\ref{eq56}).
Rearranging terms and using (\ref{eq54}) and (\ref{eq55}), this
equation can be rewritten as
\begin{equation}
 S_z = - \frac{\alpha \beta \sin \phi}{(\alpha^2 + \beta^2) k}
  \left( \frac{\partial S_y}{\partial x}
    - \frac{\partial S_x}{\partial y} \right)
 = \left[- \frac{\alpha \beta \sin \phi}{(\alpha^2 + \beta^2) k} \right]
  \vec{\zeta} \cdot \hat{\bf z} ,
 \label{eq61}
\end{equation}
where
\begin{equation}
 \vec{\zeta} \equiv \left(
  \begin{array}{ccc}
   \hat{\bf x} & \hat{\bf y} & \hat{\bf z} \\
   \displaystyle \frac{\partial}{\partial x} &
   \displaystyle \frac{\partial}{\partial y} & 0 \\
   S_x & S_y & 0
  \end{array} \right) .
 \label{eq62}
\end{equation}
According to (\ref{eq61}), the presence of a polarization state gives
rise to a flow along the $z$-direction in terms of the vorticity
manifested by the fields $S_x$ and $S_y$, which may lead the photon
trajectories to display loops out of the $XY$ plane.
Nodal structures and other singularities and topological structures
may then appear, as shown by Nye \cite{nye2} within the context of wave
dislocations \cite{nye1} and by other authors within the context of the
Riemann-Silberstein complex formulation of Maxwell's equations
\cite{berry,bialynicki5,kaiser} (see Appendix~\ref{appA}).
Experimentally, what one would observe on the $XZ$ plane is simply the
typical fringe-like interference pattern constituted by dark and light
parallel strips, which results from the accumulation of photons
arriving at this plane.
Note that (\ref{eq57}) describes the interference pattern, which
results from transporting the EME density through space (from the slits
to some detection screen) in accordance to the guidance or continuity
equation (\ref{eq44bb}).
This means that, if we make a histogram with the arrivals of a
statistical distribution of photon trajectories along the $x$-direction
(from now on, we will label the normalized amount of such counts as
$\Sigma_i$, with $i=x,z$), the well-known interference pattern emerges,
as can be seen in Fig.~\ref{fig2}(a).
However, from (\ref{eq62}), all the arrivals at a certain height $z_f$
will not arise from positions at the slits at the same height $z_0=z_f$,
but there is a flux upwards and downwards which breaks the longitudinal
(along the $z$-direction) symmetry of the experiment when it is studied
from the viewpoint of photon trajectories.
This gives rise to a certain distribution of arrivals along the
$z$-direction, as shown in Fig.~\ref{fig2}(b).
Since the photon trajectories distribute evenly around $z_0$ (here,
along positive and negative $z$, since we have chosen $z_0 = 0$), as
can be appreciated in Fig.~\ref{fig1}(d), their distribution is also
going to be symmetric with respect to $z_0$ in the histogram of
Fig.~\ref{fig2}(b).
Nonetheless, we would like to point out that this effect along the
$z$-axis arises because we are considering a particular initial value
of the $z$-coordinate.
If this coordinate is sampled properly up and down, this effect cannot
be appreciated, but only a homogenous distribution, unlike what happens
along the $x$-coordinate.

\section{\label{sec6} The EM field and photon trajectories behind two
slits, each followed by a linear polarizer}

According to the Arago-Fresnel laws \cite{barakat}, if two diffracted
beams with different polarization states interfere, the visibility of
the interference pattern will decrease.
Indeed, if the polarization states are orthogonal the pattern will
disappear totally.
In order to describe this effect with photon trajectories, we consider
the EM field behind a grating with two slits, each followed by a linear
polarizer, such that behind the slit 1 there is a polarizer with its
polarization axis oriented along the $z$-axis and behind the slit 2
the polarizer is oriented along the $x$-axis.
As in Sec.~\ref{sec4}, we will also start expressing the total EM field
in terms of the superposition of two fields, each propagating from one
of the slits (see (\ref{eqn65}) and (\ref{eqn66})).
However, due to the presence of the polarizers and their filtering
effect on the incident EM field, instead of Eqs.~(\ref {eq27}) and
(\ref {eq28}), now we will express the electric and magnetic fields
in terms of $\psi_1$ and $\psi_2$, described by (\ref {eqsu1}), i.e.,
\begin{eqnarray}
 {\bf E}_e = \alpha \psi_1({\bf r}) \hat{\bf z} ,
 \label{eqn72} \\
 {\bf H}_h = \beta e^{i\phi} \sqrt{\frac{\epsilon_0}{\mu_0}} \
   \psi_2({\bf r}) \hat{\bf z} .
 \label{eqn73}
\end{eqnarray}
By substituting (\ref{eqn72}) and (\ref{eqn73}) into (\ref{eq25}) and
(\ref{eq26}), we will obtain the EM field behind the two slits covered
by the polarizers,
\begin{eqnarray}
 {\bf E} & = &
  \frac{i \beta e^{i\phi}}{k}
    \frac{\partial \psi_2}{\partial y} \ \hat{\bf x}
  - \frac{i \beta e^{i\phi}}{k}
    \frac{\partial \psi_2}{\partial x} \ \hat{\bf y}
  + \alpha \psi_1 \hat{\bf z} ,
 \label{eqn74}\\
 {\bf H} & = &
   -\frac{i \alpha}{\omega \mu_0}
    \frac{\partial \psi_1}{\partial y} \ \hat{\bf x}
  + \frac{i \alpha}{\omega \mu_0}
    \frac{\partial \psi_1}{\partial x} \ \hat{\bf y}
  + \frac{k \beta e^{i\phi}}{\omega \mu_0} \ \psi_2 \hat{\bf z} .
 \label{eqn75}
\end{eqnarray}
From these fields, we obtain the expressions for the EME density and
the Poynting vector,
\begin{eqnarray}
 U & = & \frac{\alpha^2}{4\omega^2\mu_0} \left(
   \frac{\partial \psi_1}{\partial x}
     \frac{\partial \psi_1^*}{\partial x}
 + \frac{\partial \psi_1}{\partial y}
     \frac{\partial \psi_1^*}{\partial y}
 + k^2 \psi_1 \psi_1^* \right)
 \nonumber \\
 & & + \frac{\beta^2}{4\omega^2\mu_0} \left(
   \frac{\partial \psi_2}{\partial x}
     \frac{\partial \psi_2^*}{\partial x}
 + \frac{\partial \psi_2}{\partial y}
     \frac{\partial \psi_2^*}{\partial y}
 + k^2 \psi_2 \psi_2^* \right)\nonumber\\
 & = & U_1 + U_2
 \label{eq67b}
\end{eqnarray}
and
\begin{eqnarray}
 S_x & = & \frac{i\alpha^2}{4\omega\mu_0}
  \left( \psi_1 \frac{\partial \psi_1^*}{\partial x}
  - \psi_1^* \frac{\partial \psi_1}{\partial x} \right)
  + \frac{i\beta^2}{4\omega\mu_0}
  \left( \psi_2 \frac{\partial \psi_2^*}{\partial x}
  - \psi_2^* \frac{\partial \psi_2}{\partial x} \right) ,
 \label{eq70b} \\
 S_y & = & \frac{i\alpha^2}{4\omega\mu_0}
  \left( \psi_1 \frac{\partial \psi_1^*}{\partial y}
  - \psi_1^* \frac{\partial \psi_1}{\partial y} \right)
  + \frac{i\beta^2}{4\omega\mu_0}
  \left( \psi_2 \frac{\partial \psi_2^*}{\partial y}
  - \psi_2^* \frac{\partial \psi_2}{\partial y} \right) ,
 \label{eq71b} \\
 S_z & = &
  \frac{\alpha\beta e^{i\phi}}{4k\omega\mu_0}
  \left( \frac{\partial \psi_2}{\partial y}
     \frac{\partial \psi_1^*}{\partial x}
  - \frac{\partial \psi_2}{\partial x}
     \frac{\partial \psi_1^*}{\partial y} \right)
 \nonumber \\
  & & - \frac{\alpha\beta e^{-i\phi}}{4k\omega\mu_0}
  \left( \frac{\partial \psi_1}{\partial y}
     \frac{\partial \psi_2^*}{\partial x}
  - \frac{\partial \psi_1}{\partial x}
     \frac{\partial \psi_2^*}{\partial y} \right) ,
 \label{eq72b}
\end{eqnarray}
respectively.
As can be noticed in (\ref{eq67b}), $U$ describes the bare addition of
EME densities associated with the fields diffracted by each slit, with
no interference term.
This means that no interference pattern is going to be observed at
the detection screen, in accordance to the Arago-Fresnel law for
the interference of two beams with orthogonal polarization states
(perpendicularly polarized in the case of linear polarization
\cite{hunt} or with opposite handedness in the case of elliptic or
circular polarization \cite{pescetti}).
Regarding the Poynting vector, we note that the EME flux along the $x$
and $y$-direction is also given by the simple addition of fluxes coming
from each slit.
Thus, unlike the diffraction problem dealt with in the previous
section, now it is not possible to factorize the flux (neither the
EME density) in terms of its spatial and polarization parts.
This has a consequence on the EME flux along the $z$-direction, which
does not satisfy the rotational character described by (\ref{eq62}).
On the other hand, since the third and forth terms in (\ref{eq72b}) do
not vanish, there will be an EME flux along the $z$-direction even in
the case of linear polarization (remember from Sec.~\ref{sec5} that,
for linear polarization, $S_z=0$).
This can also be seen from the photon-trajectory equations,
\begin{eqnarray}
 \frac{dz}{dx} & = & \frac{\displaystyle
   \frac{i \alpha \beta e^{i\phi}}{k}
   \left( \frac{\partial \psi_2}{\partial x}
          \frac{\partial \psi_1^*}{\partial y}
        - \frac{\partial \psi_2}{\partial y}
          \frac{\partial \psi_1^*}{\partial x} \right)}
   {\displaystyle
   \alpha^2 \left( \psi_1 \frac{\partial \psi_1^*}{\partial y}
  - \psi_1^* \frac{\partial \psi_1}{\partial y} \right)
 + \beta^2 \left( \psi_2 \frac{\partial \psi_2^*}{\partial y}
  - \psi_2^* \frac{\partial \psi_2}{\partial y} \right)}
 \nonumber \\
 & & - \frac{\displaystyle \frac{i \alpha \beta e^{-i\phi}}{k}
   \left( \frac{\partial \psi_1}{\partial x}
          \frac{\partial \psi_2^*}{\partial y}
        - \frac{\partial \psi_1}{\partial y}
          \frac{\partial \psi_2^*}{\partial x} \right)}
   {\displaystyle
   \alpha^2 \left( \psi_1 \frac{\partial \psi_1^*}{\partial y}
  - \psi_1^* \frac{\partial \psi_1}{\partial y} \right)
 + \beta^2 \left( \psi_2 \frac{\partial \psi_2^*}{\partial y}
  - \psi_2^* \frac{\partial \psi_2}{\partial y} \right)} ,
 \label{eq79} \\
 \frac{dy}{dx} & = & \frac{\displaystyle
   \alpha^2 \left( \psi_1 \frac{\partial \psi_1^*}{\partial y}
  - \psi_1^* \frac{\partial \psi_1}{\partial y} \right)
 + \beta^2 \left( \psi_2 \frac{\partial \psi_2^*}{\partial y}
  - \psi_2^* \frac{\partial \psi_2}{\partial y} \right)}
   {\displaystyle
   \alpha^2 \left( \psi_1 \frac{\partial \psi_1^*}{\partial x}
  - \psi_1^* \frac{\partial \psi_1}{\partial x} \right)
 + \beta^2 \left( \psi_2 \frac{\partial \psi_2^*}{\partial x}
  - \psi_2^* \frac{\partial \psi_2}{\partial x} \right)} .
 \label{eq79b}
\end{eqnarray}
Note in (\ref{eq79}) that $dz/dx$, effectively, does not vanish, not
only for $\phi =0$ or $\pi$ (the conditions for linear polarization),
but neither for any other $\phi$ value.

\begin{figure}
 \begin{center}
 \includegraphics[width=7.5cm]{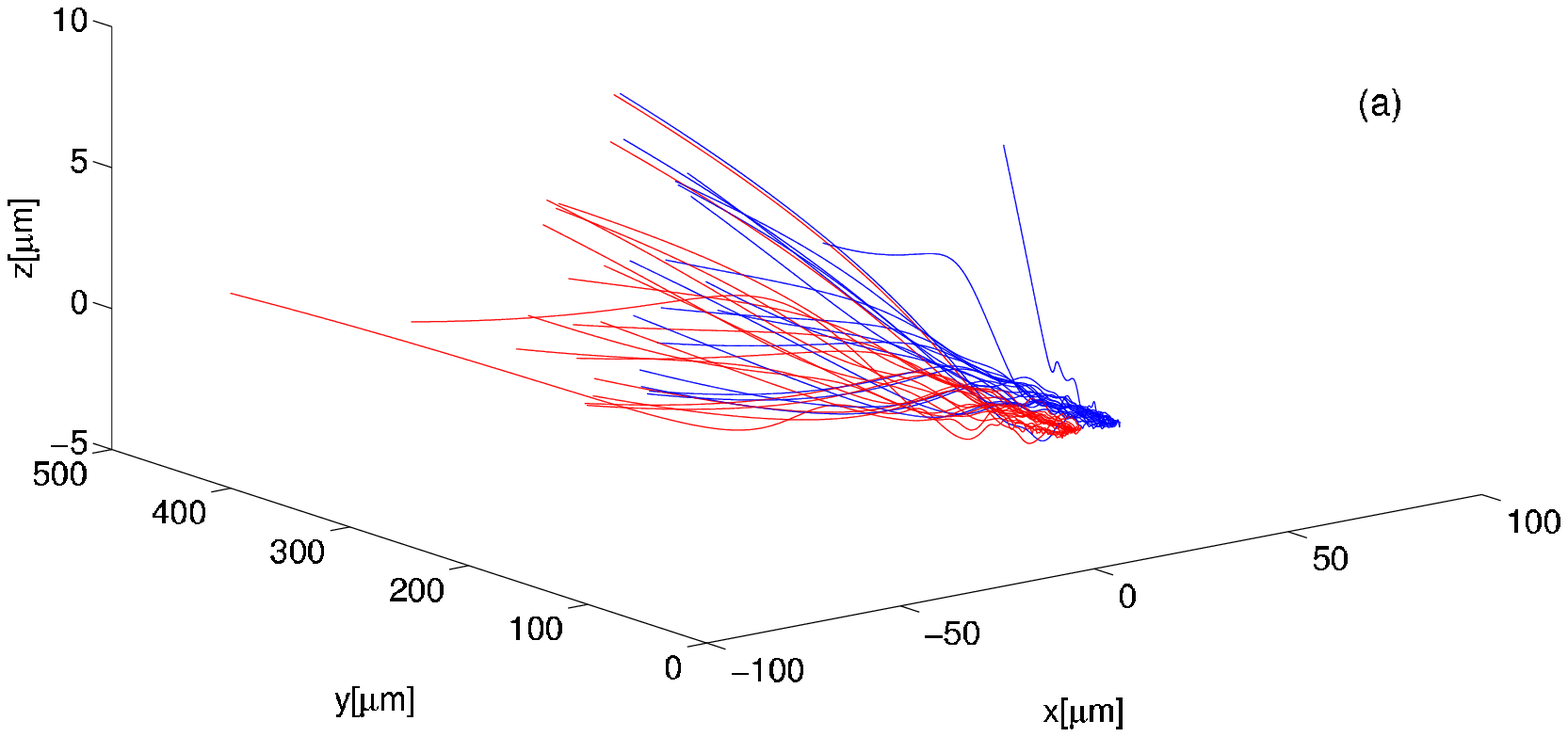}
 \includegraphics[width=7.5cm]{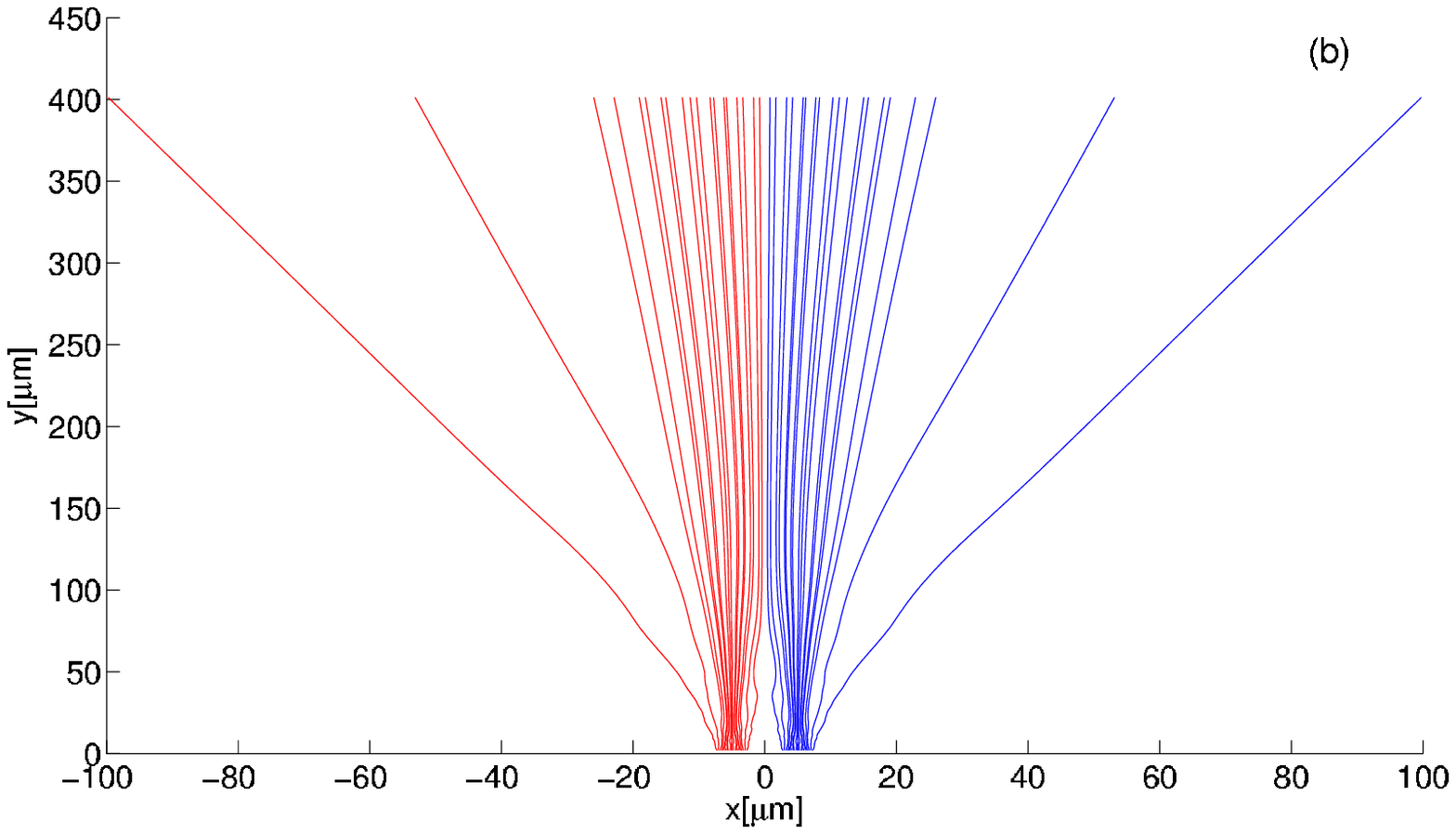}
 \includegraphics[width=7.5cm]{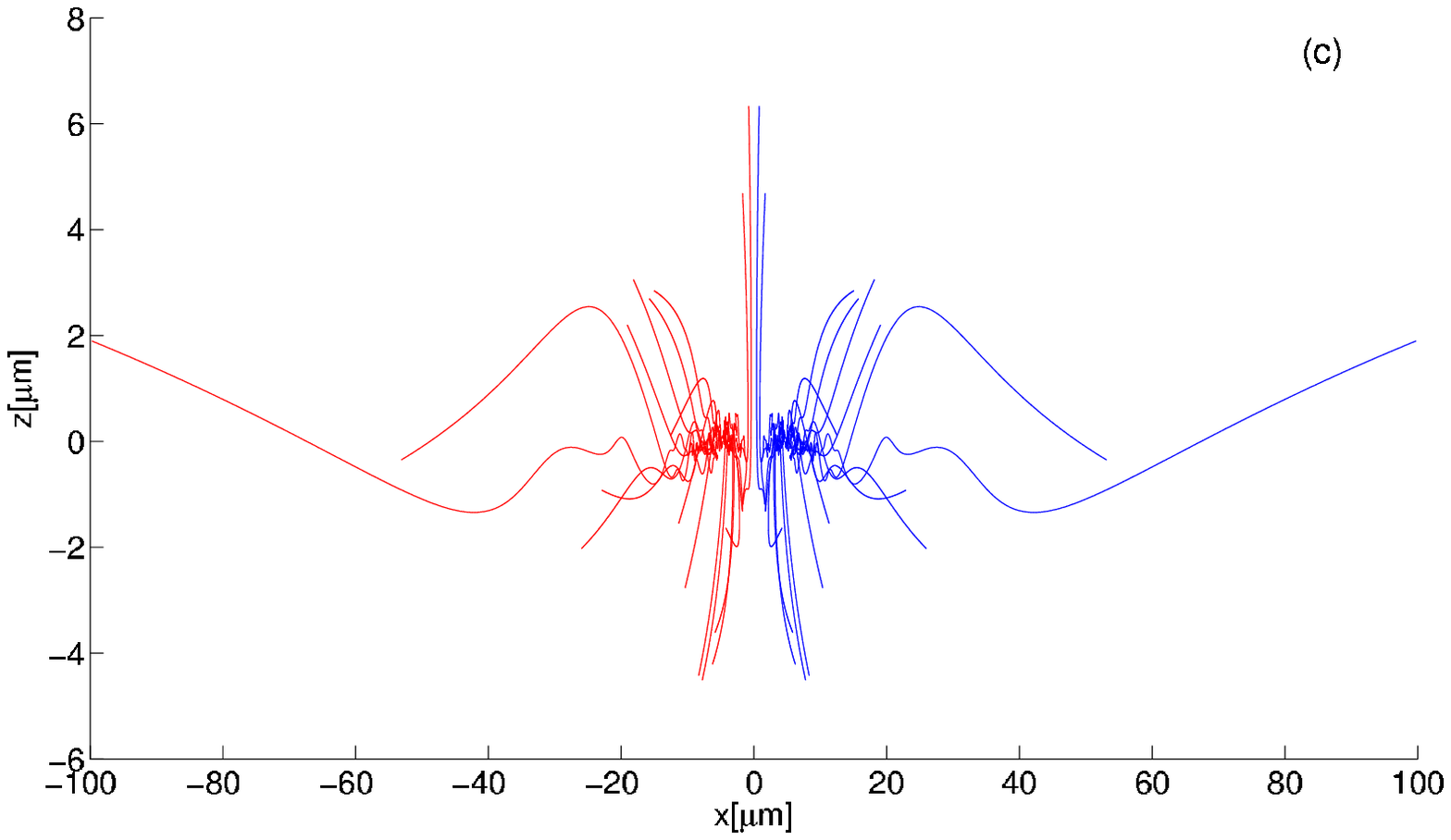}
 \includegraphics[width=7.5cm]{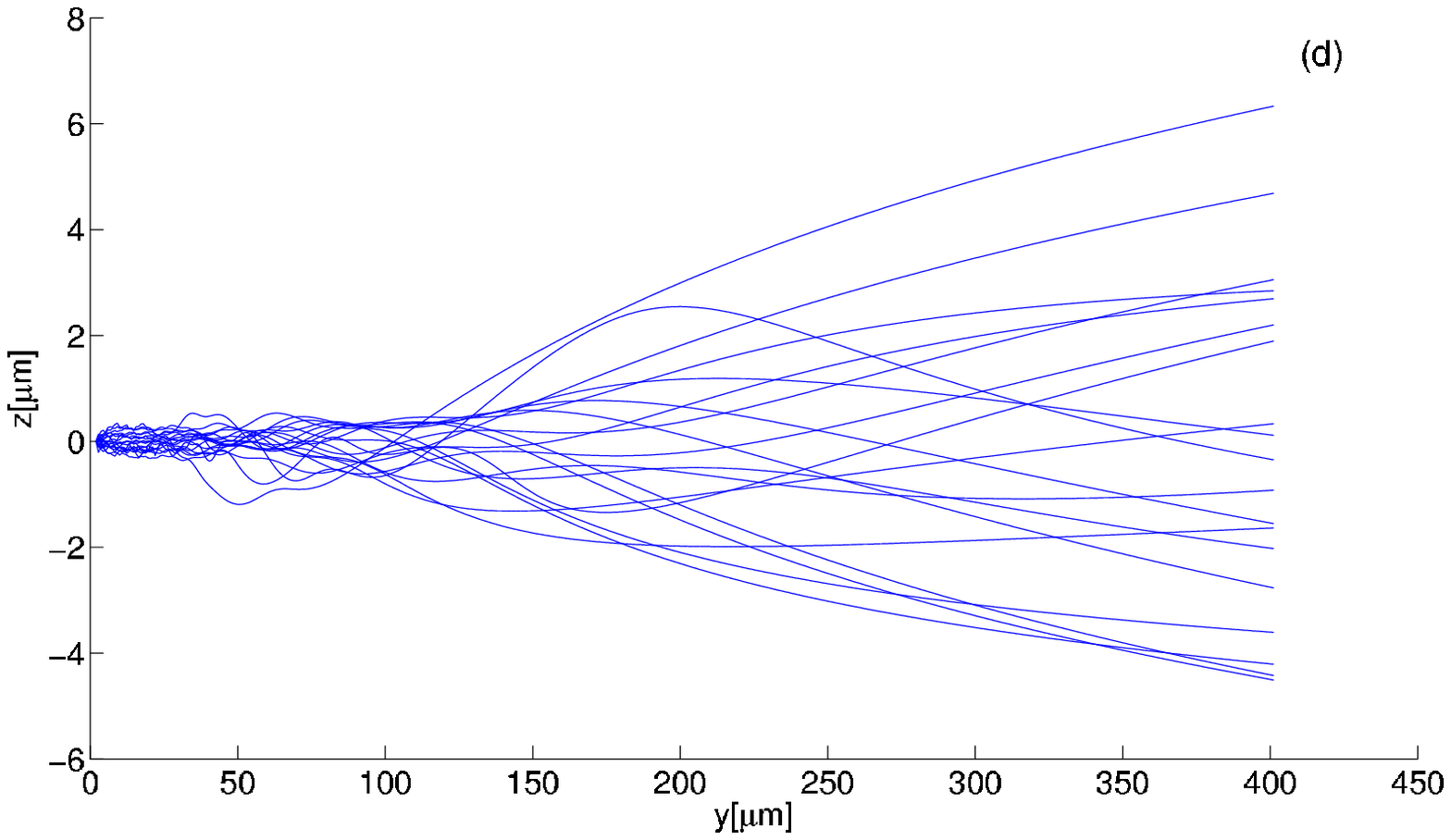}
 \caption{\label{fig3}
  Photon trajectories (15 for each slit) behind a two-slit grating,
  where each slit is followed by polarizers with orthogonal axes and
  the incident EM field is linearly polarized ($\alpha = \beta$,
  $\phi=0$): (a) 3D view, (b) $XY$ projection, (c) $XZ$ projection
  and (d) $YZ$ projection.
  Because the incident EM field is linearly polarized, the ensembles
  leaving each slit behave exactly the same and, therefore, the
  trajectories exiting through slit 1 look the same as those exiting
  through slit 2.
  The parameters considered in the simulation are: $\lambda = 500$~nm,
  $d = 20 \lambda = 10$~$\mu$m and $\delta = d/2 = 5$~$\mu$m.}
 \end{center}
\end{figure}

\begin{figure}
 \begin{center}
 \includegraphics[width=7.5cm]{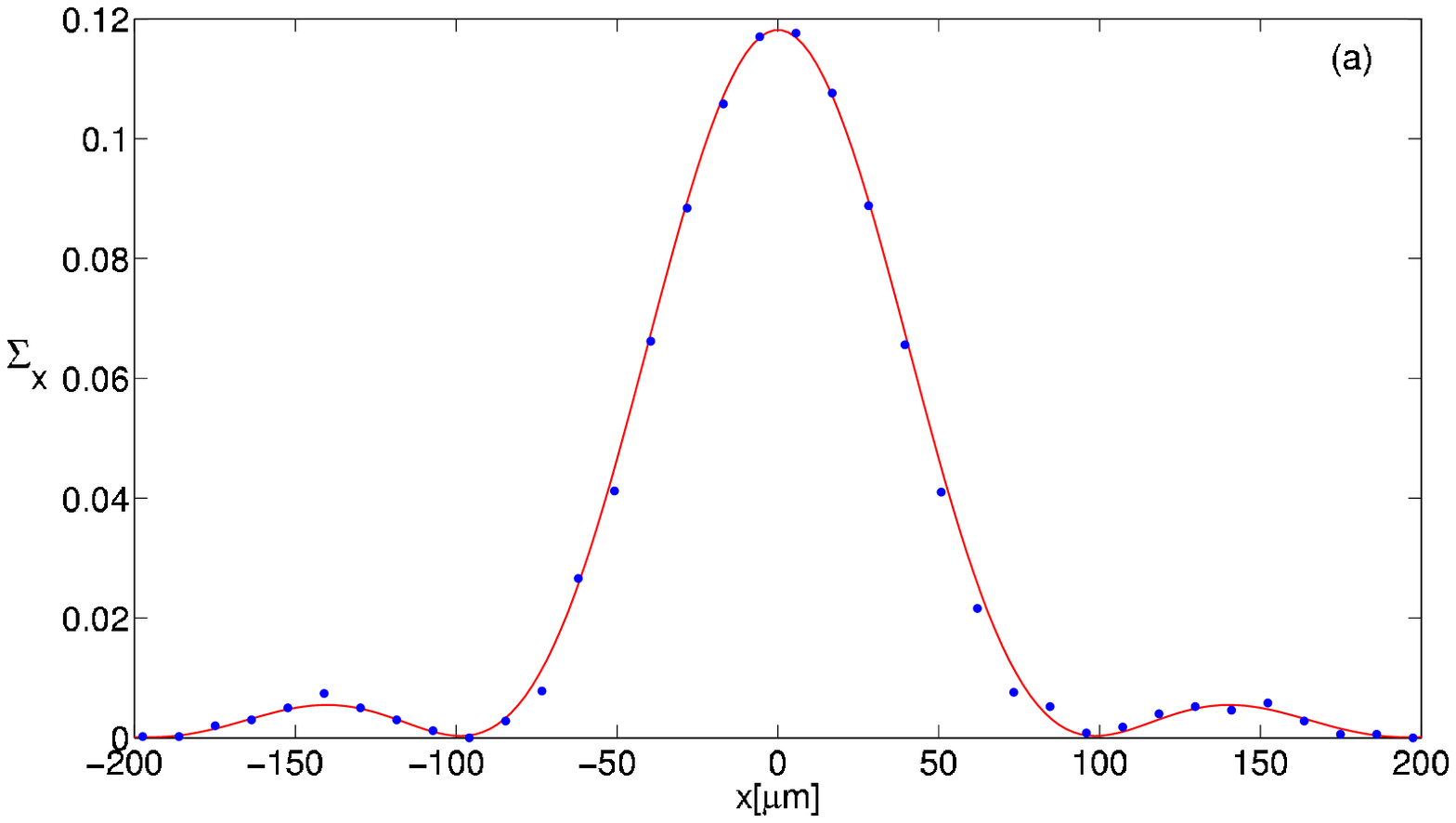}
 \includegraphics[width=7.5cm]{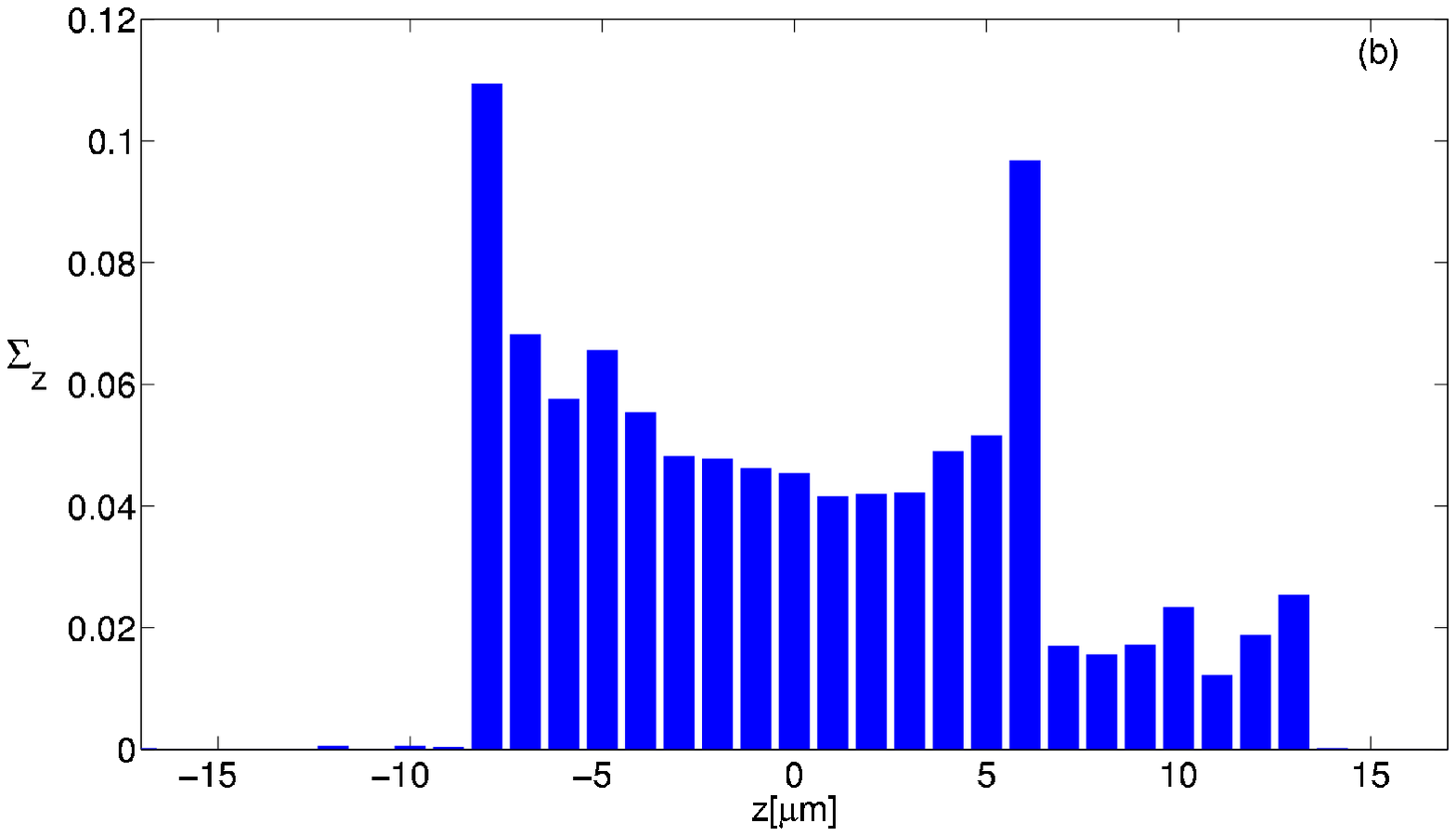}
 \caption{\label{fig4}
  Histograms built up by counting the end points of individual photon
  trajectories associated with the incident EM field linearly polarized
  ($\alpha = \beta$, $\phi=0$) and where, as in Fig.~\ref{fig3}, each
  slit is followed by polarizers with orthogonal axes.
  The detection screen is at $L = 1$~mm from a two-slit grating and the
  histograms represent counts: (a) along the $x$-direction and (b)
  along the $z$-direction.
  In part (a), the red solid line indicates the theoretical curve
  predicted by standard Electromagnetism, according to (\ref{eq67b}).
  The parameters considered in the simulation are: $\lambda = 500$~nm,
  $d = 20 \lambda = 10$~$\mu$m and $\delta = d/2 = 5$~$\mu$m.
  The total number of trajectories considered is 5,000, with initial
  conditions homogenously distributed along each slit.}
 \end{center}
\end{figure}

In Fig.~\ref{fig3}, we observe an ensemble of photon trajectories for
an incident EM field linearly polarized ($\alpha = \beta$, $\phi=0$).
As before, the parameters considered in the simulation are: $\lambda =
500$~nm, $d = 20 \lambda = 10$~$\mu$m and $\delta = d/2 = 5$~$\mu$m.
As can be noticed, when polarizers with orthogonal polarization
directions act on the diffracted wave, the topology of the flow
lines changes dramatically when compared with that observed in
Fig.~\ref{fig1}.
When one looks at the $XY$ projection (see Fig.~\ref{fig3}(b)), the
wiggling behavior that gives rise to the different interference fringes
of the pattern in the Fraunhofer region are lacking.
This reflects the fact that the EME density, described by (\ref{eq67b}),
is just the sum of the EME densities associated with the components
of the wave that arise from each slit, which can be appreciated in the
histogram presented in Fig.~\ref{fig4}(a).
This histogram reproduces the distribution pattern of the photon
trajectories in the Fraunhofer region, which simply consists of the
direct sum of the two single-slit diffraction patterns associated with
the EM field arising from each slit.
In particular, since the grating is assumed as fully transparent along
the space covered by the slits and opaque elsewhere, the single-slit
diffraction pattern is a square sinc function, this leading to the
appearance of the maxima and minima observed in Fig.~\ref{fig4}(a).
The reason why this pattern looks like the one arising from a single
slit rather than the sum of two of them is very simple.
The observation distance $L$ (1~mm) is much larger than the distance
between the two slits $d$ (10~$\mu$m) and, therefore, the single-slit
diffraction pattern associated with each slit basically overlap.
For example, the first minima for slit 1 occur at $x_{\rm min}^{1,-}
\approx -105$~$\mu$m and $x_{\rm min}^{1,+} \approx 95$~$\mu$m,
while for slit 2 they are at $x_{\rm min}^{2,-} \approx -95$~$\mu$m
and $x_{\rm min}^{2,+} \approx 105$~$\mu$m; therefore, the minima of
the total pattern will appear at $x_{\rm min}^\pm \approx \pm
100$~$\mu$m, as can be seen in Fig.~\ref{fig4}(a).
It is worth mentioning that the loss of the interference fringes as
well as the lack of wiggling features in the trajectories is similar to
the behavior found in trajectories for massive particles in the case of
decoherence \cite{asanz-EPJD,asanz-CPL}.

If we look at the topology of the photon trajectories along the
$z$-direction (see Figs.~\ref{fig3}(c) and (d)), we note that there
is a symmetry with respect to $x = 0$ (see Fig.~\ref{fig3}(c)).
This is the result of the breaking of the rotationality property
mentioned in the previous section.
The same ``symmetry'' breaking can also be observed by looking at the
distribution of the trajectories along the $z$-direction, as shown in
Fig.~\ref{fig4}(b).
If we changed the initial polarization state, from $\phi = 0$ to $\phi
= \pi$, we would obtain the same pattern, but inverted with respect to
$z = 0$.
Again, as we discussed in Sec.~\ref{sec5}, this is an effect arising
from considering a particular initial value of the $z$-coordinate for
the photon trajectories, which will disappear when a sampling along
the $z$-axis is considered.

\begin{figure}
 \begin{center}
 \includegraphics[width=7.5cm]{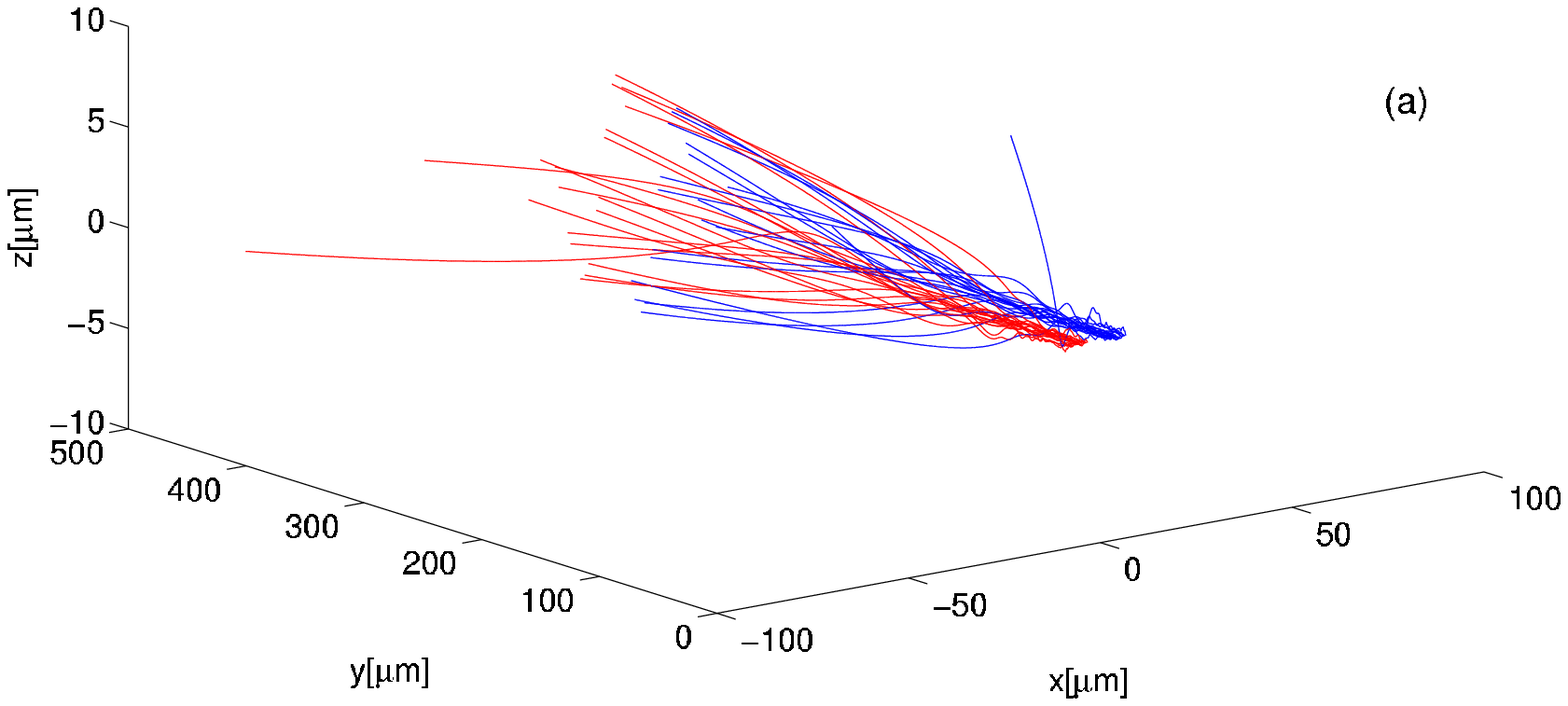}
 \includegraphics[width=7.5cm]{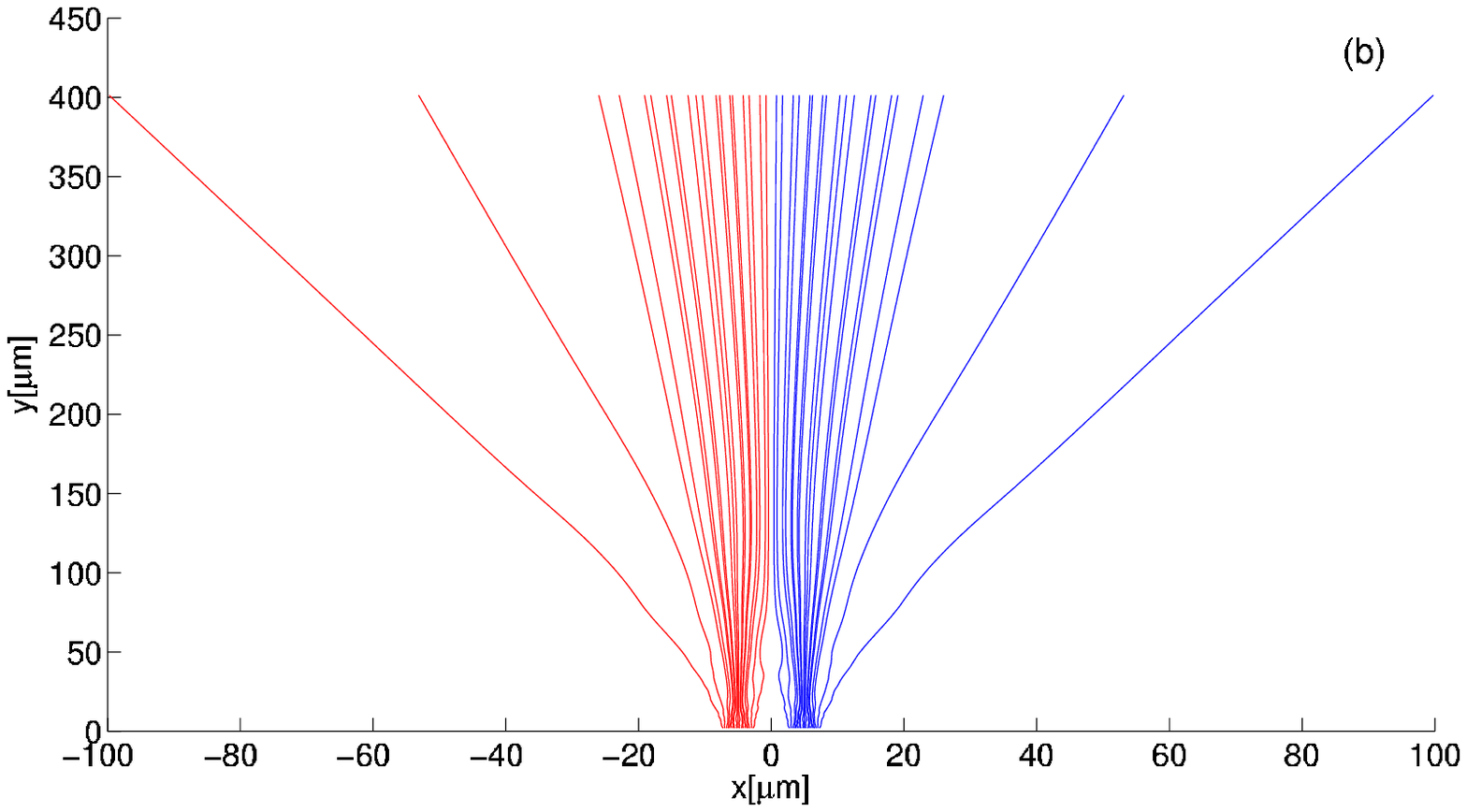}
 \includegraphics[width=7.5cm]{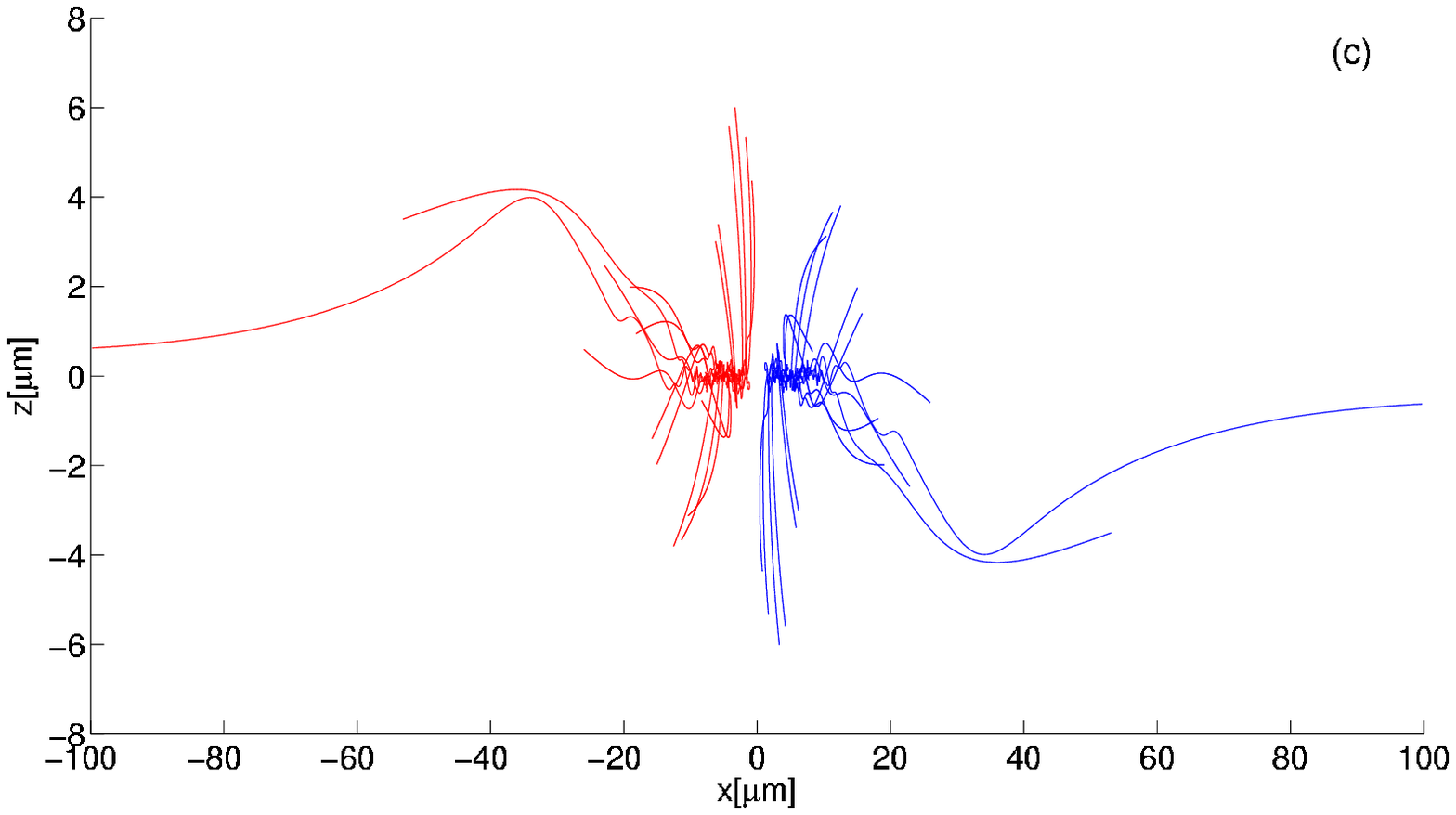}
 \includegraphics[width=7.5cm]{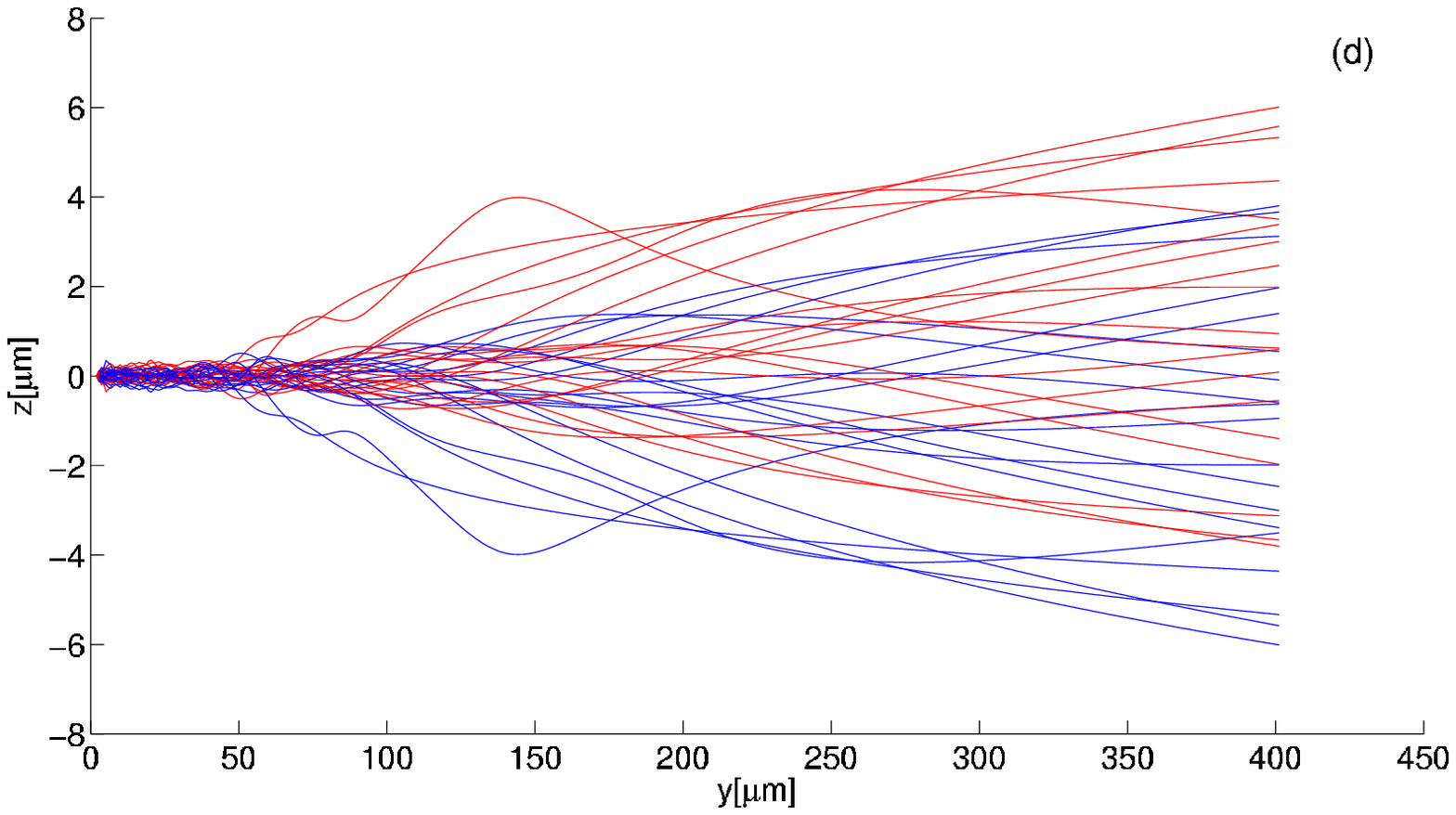}
 \caption{\label{fig6}
  Photon trajectories (15 for each slit) behind a two-slit grating,
  where each slit is followed by polarizers with orthogonal axes and
  the incident EM field is circularly polarized ($\alpha = \beta$,
  $\phi = \pi/2$): (a) 3D view, (b) $XY$ projection, (c) $XZ$
  projection and (d) $YZ$ projection.
  The parameters considered in the simulation are: $\lambda = 500$~nm,
  $d = 20 \lambda = 10$~$\mu$m and $\delta = d/2 = 5$~$\mu$m.}
 \end{center}
\end{figure}

In the case of an incident EM field circularly polarized, displayed
in Fig.~\ref{fig6}, we find that the photon-trajectory projections on
the $XY$ plane (see Fig.~\ref{fig6}(b)) look exactly the same as those
for the incident EM field linearly polarized (see Fig.~\ref{fig3}(b)),
in accordance to Eq.~(\ref{eq79b}).
However, as seen in Sec.~\ref{sec5}, due to the rotationality
associated with circular polarization, a breaking of the specular
symmetry with respect to $x = 0$ will take place, and the EME flux
leaving each slit is going to be different, as shown in
Figs.~\ref{fig6}(c) and \ref{fig6}(d).
This will give rise to a histogram like the one shown in
Fig.~\ref{fig4}(a) when photon are counted along the $x$-direction
and another similar to that of Fig.~\ref{fig2}(b) when they are
counted along the $z$-direction.


\section{Final discussion and conclusions}
 \label{sec7}

Interference experiments with non-relativistic massive particles and
photons can be well understood on the theoretical grounds provided by
the (non-relativistic) Schr\"odinger and Maxwell equations,
respectively.
Thus, if one wishes to determine photon trajectories on the same
footing as Bohmian trajectories, the most appropriate theoretical
framework is the one based on Maxwell's equations, more specifically
their hydrodynamical formulation, proposed by Bialynicki-Birula
\cite{bialynicki1,bialynicki2,bialynicki3,bialynicki4}.
Within this formulation, a close analogy can be established between
the trajectory equations based on the latter and Bohm's approach when
the particle aspect of both light and matter is taken into
consideration.
This allows to compare on the same grounds the (Bohmian) trajectories
for massive particles with the trajectories derived for photons from
classical electromagnetism.
The photon trajectories (EME flow lines) are determined from the
Poynting vector, with the components of the electric and magnetic
vector fields expressed in terms of a function that explicitly takes
into account the boundary conditions imposed by the grating.
It is remarkable that, in the case of photons, any trajectory-based
interpretation will be complementary to the standard Huyghens' one,
based on the superposition of secondary wavelets.
Furthermore, the topology displayed by the photon trajectories is
strikingly similar to that displayed by massive particles.
As happens in the case of massive particles \cite{sanz3}, such a
topology can also be inferred and explained from the corresponding
trajectory equation.

Here, we have considered the Bialynicki-Birula hydrodynamical framework
to analyze the effects of polarization on interference pattern for
waves which arise from an initially polarized monochromatic EM plane
wave.
According to the Arago-Fresnel laws of interference with polarized
light in Young-type experiments \cite{barakat}, if, for example, the
two interfering beams are linearly polarized in orthogonal directions
\cite{hunt} or both are elliptically polarized but with opposite
handedness \cite{pescetti}, no interference will be observed.
The photon trajectories presented in this work show how the EME density
distributes in configuration space, giving rise to the appearance or
not of interference fringes.
This thus constitutes a very interesting corpuscular-like description
of the Arago-Fresnel laws for the interference of polarized light,
consistent with the more standard wave description.
Moreover, it is clear that the photon wave function can be interpreted
as an energy localization amplitude or detection amplitude within the
hydrodynamical approach to Electromagnetism.
Now, from the agreement between the histograms and the corresponding
EME density distribution (which follows from (\ref{eq44bb}) or
(\ref{eq44})), this density can also be understood as a probability
density function in the sense that arrivals are going to distribute
statistically according to it, as can be seen in the recent experiment
carried out by Dimitrova and Weis \cite{weis}, for example.


\section*{Acknowledgements}

MD and MB acknowledge support from the Ministry of Science of Serbia
under Project ``Quantum and Optical Interferometry'', No.~141003;
ASS and SMA acknowledge support from the Ministerio de Ciencia e
Innovaci\'on (Spain) under Project FIS2007-62006.
ASS also thanks the Consejo Superior de Investigaciones Cient\'{\i}ficas
(Spain) for a JAE-Doc Contract.


\appendix

\section{Appendix A: Photon trajectories within the Riemann-Silberstein
formulation of Electromagnetism}
\label{appA}

Both Maxwell's and Schr\"odinger's equations describe the evolution
of fields in configuration space, namely the EM field and probability
fields associated with non-relativistic massive particles,
respectively.
From these fields, one can obtain trajectories which show how they
evolve in space, namely photon trajectories or Bohmian trajectories,
respectively.
However, Maxwell's equations look quite different formally from the
Schr\"odinger equation as they are usually given.
A formulation that allows one to put Maxwell's equations on similar
formal grounds as Schr\"odinger's one (but without $\hbar$ due to the
lack of mass of photons) is the complex form of Maxwell's equations
\cite{riemann,bateman}, which is based on the so-called
Riemann-Silberstein complex EM vector \cite{bialynicki1,bialynicki2},
\be
 \tilde{\bf F}({\bf r},t) = \frac{1}{\sqrt{2}}
  \left[ \sqrt{\epsilon_0} \ \! \tilde{\bf E}({\bf r},t)
   + i \sqrt{\mu_0} \ \! \tilde{\bf H}({\bf r},t) \right] ,
 \label{rsvector}
\ee
with the change of variable
\ba
 \tilde{\bf E}({\bf r},t) & = & \frac{1}{\sqrt{2\epsilon_0}}
  \left( \tilde{\bf F} + \tilde{\bf F}^* \right) ,
 \label{ef} \\
 \tilde{\bf H}({\bf r},t) & = & \frac{1}{i\sqrt{2\mu_0}}
  \left( \tilde{\bf F} - \tilde{\bf F}^* \right) ,
 \label{bf}
\ea
where $\tilde{\bf E}$ and $\tilde{\bf H}$ are real fields.
Introducing (\ref{ef}) and (\ref{bf}) into Maxwell's equations in the
absence of electrical charge densities, we obtain
\ba
 i\ \frac{\partial \tilde{\bf F}}{\partial t}
  & = & c \nabla \times \tilde{\bf F}
 \label{feq1} \\
 \nabla \cdot \tilde{\bf F} & = & 0 .
 \label{feq12}
\ea
As can be noticed, (\ref{feq1}) is the analog for photons of the Dirac
equation for massive particles, while (\ref{feq12}) describes the
conservation of the EME density through space.
This analogy within the Riemann-Silberstein formulation becomes more
apparent by gathering (\ref{feq1}) and (\ref{feq12}) in a single
equation.
This is done by applying the operator $-i\partial/\partial t$ to both
sides of (\ref{feq1}) and then rearranging terms taking into account
(\ref{feq12}) and the vectorial relation
\be
 \nabla \times (\nabla \times {\bf A}) =
  \nabla (\nabla \cdot {\bf A}) - \nabla^2 {\bf A} ,
\ee
where ${\bf A}$ is a general vector field.
This renders
\be
 \frac{\partial^2 \tilde{\bf F}}{\partial t^2} =
  c^2 \nabla^2 \tilde{\bf F} ,
 \label{feq2}
\ee
which has the well-known form of the Klein-Gordon equation.
In the particular case of diffraction problems, which can be reduced
to boundary condition problems because of their time-independence, the
space and time parts of (\ref{feq2}) are separable.
Thus, $\tilde{\bf F}_i$ can be decomposed as
\be
 \tilde{\bf F}_i({\bf r},t) = {\bf F}_i({\bf r}) \ \! \phi_i(t)
   = {\bf F}_i({\bf r}) \ \! e^{-i\omega t} ,
 \label{e4}
\ee
where the space part, ${\bf F}_i$, satisfies Helmholtz's equation,
\be
 \nabla^2 {\bf F}_i({\bf r}) + k^2 {\bf F}_i({\bf r}) = 0 ,
 \label{e2}
\ee
and its time-dependent part the differential equation
\be
  \frac{\partial^2 \phi_i(t)}{\partial t^2} = -\omega^2 \phi_i(t) ,
 \label{e3}
\ee
with $c = \omega/k$.

Within this formulation, the EME density is given by
\be
 \mathcal{U} = \frac{1}{2}
  \left( \epsilon_0 \ \! \tilde{\bf E} \cdot \tilde{\bf E}
  + \mu_0 \ \! \tilde{\bf H} \cdot \tilde{\bf H} \right)
  = \tilde{\bf F} \cdot \tilde{\bf F}^*
 \label{fu-r}
\ee
and its flux, described by the {\it Poynting vector}, is
straightforwardly obtained after developing the time-derivative
of $\mathcal{U}$ to yield
\be
 \mathcal{S} = \tilde{\bf E} \times \tilde{\bf H}
  = i c \tilde{\bf F} \times \tilde{\bf F}^* ,
 \label{fp-r}
\ee
where the relation
\be
 \nabla ({\bf A} \times {\bf B}) =
  {\bf B} \cdot \nabla \times {\bf A} -
   {\bf A} \cdot \nabla \times {\bf B}
 \label{nablarel}
\ee
for any two vectors, ${\bf A}$ and ${\bf B}$, has been taken into
account.
From these magnitudes, we can now obtain the stationary photon
trajectories within the Riemann-Silberstein formulation as,
\be
 \frac{d {\bf r}}{ds} = \frac{1}{c} \frac{\bf S}{U}
  = i \frac{\langle \tilde{\bf F} \times \tilde{\bf F}^* \rangle}
   {\langle \tilde{\bf F} \cdot \tilde{\bf F}^* \rangle}
 \label{eqb8b}
\ee
(here, $\langle \tilde{\bf A} \rangle$ denotes the time-average of
the magnitude $\tilde{\bf A}$), which transport the time-averaged EME
density, $U$, as described by the (also time-averaged) Poynting vector,
${\bf S}$.

If the electric and magnetic fields are complex and the definition of
the Riemann-Silberstein vector is kept as in (\ref{rsvector}), i.e.,
the real and imaginary parts are given by the electric and magnetic
fields, respectively, then we need to include into the formulation
two of these vectors in order to have a complete description of the
problem, each one associated with the real or the imaginary parts of
the electric and magnetic fields.
That is, if
\be
 \tilde{\bf E} = \tilde{\bf E}_1 + i \tilde{\bf E}_2 , \qquad
 \tilde{\bf H} = \tilde{\bf H}_1 + i \tilde{\bf H}_2 ,
\ee
with $\tilde{\bf E}_i$ and $\tilde{\bf H}_i$ ($i = 1,2$) being real
vector fields satisfying the corresponding Maxwell equations, we will
have
\ba
 \tilde{\bf F}_1 & = & \frac{1}{\sqrt{2}}
  \left( \sqrt{\epsilon_0} \ \! \tilde{\bf E}_1
   + i \sqrt{\mu_0} \ \! \tilde{\bf H}_1  \right) ,
 \nonumber \\
 \tilde{\bf F}_2 & = & \frac{1}{\sqrt{2}}
  \left( \sqrt{\epsilon_0} \ \! \tilde{\bf E}_2
   + i \sqrt{\mu_0} \ \! \tilde{\bf H}_2  \right) .
\ea
The EME density (\ref{fu-r}) and the Poynting vector (\ref{fp-r})
then read as
\ba
 \mathcal{U} & = & \frac{1}{2}
 \left( \epsilon_0 \ \! \tilde{\bf E} \cdot \tilde{\bf E}^*
 + \mu_0 \ \! \tilde{\bf H} \cdot \tilde{\bf H}^* \right)
  = \sum_{i=1,2} \tilde{\bf F}_i \cdot \tilde{\bf F}_i^* ,
 \label{fu} \\
 \mathcal{S} & = &
  {\rm Re} \left( \tilde{\bf E} \times \tilde{\bf H}^* \right)
  = {\rm Re} \left(
   i c \sum_{i=1,2} \tilde{\bf F}_i \times \tilde{\bf F}_i^* \right) ,
 \label{fp}
\ea
respectively, and their time-averaged homologous, assuming the
decomposition (\ref{e4}), as
\ba
 U & = & \frac{1}{2} \ \! \sum_{i=1,2} {\bf F}_i \cdot {\bf F}_i^* ,
  \\
 {\bf S} & = & \frac{1}{2} \ \! {\rm Re} \left( ic \ \! \sum_{i=1,2}
   {\bf F}_i \times {\bf F}_i^* \right) .
 \label{e6}
\ea

Apart from the interest of this formulation within the field of the
Fundamental Physics, it has also been considered in a more applied
way, for example, in Solid State Physics and Condensed Matter
\cite{borisov1,borisov2}.


\end{document}